\documentclass[final,1p,times]{elsarticle}

\usepackage{graphicx,psfrag,color}
\usepackage{dcolumn}
\usepackage{amsmath,amsfonts,amssymb}
\usepackage{bm}
\usepackage{mathbbol}

\newcommand{\ket}[1]{|#1\rangle}      

\usepackage{amssymb}
\usepackage{amsthm}
\usepackage{amsmath}

\journal{Annals of Physics}

\begin{document}

\begin{frontmatter}



\title{Improving the efficiency of single and multiple teleportation protocols based on the 
direct use of partially entangled states}


\author{Raphael Fortes}
\author{Gustavo Rigolin}
\ead{rigolin@ufscar.br}
\address{Departamento de F\'{i}sica,
Universidade Federal de S\~ao Carlos, S\~ao Carlos, SP 13565-905,
Brazil}

\begin{abstract}
We push the limits of the direct use of partially pure entangled states to perform quantum teleportation
by presenting several protocols in many different scenarios that achieve the optimal efficiency possible. 
We review and put in a single formalism the three major strategies known to date 
that allow one to use partially entangled states for direct quantum
teleportation (no distillation strategies permitted) and compare their efficiencies in real
world implementations. We show how one can improve the efficiency of many direct teleportation
protocols by combining these techniques. We then develop new teleportation protocols employing 
multipartite partially entangled states. The three techniques are also 
used here in order to achieve the highest efficiency possible.
Finally, we prove the upper bound for the optimal success rate for protocols based on 
partially entangled Bell states and show that some of the protocols here developed achieve such
a bound.  
\end{abstract}


\end{frontmatter}



\section{Introduction}

For many decades since its discovery,
entangled quantum states were only important in
discussions on the foundations and on the meaning of 
quantum mechanics \cite{sch35,epr,bohr,bel64}. 
By the end of the twentieth century, though, the fate
of entangled quantum states changed abruptly. 
These states are now seen and understood 
as a resource available in Nature which, 
if properly processed and handled, are capable of performing
in a very efficient way several protocols designed to process
and transmit information \cite{livrodonielsen,livrodozeilinger}.

Indeed, with the aid of entangled states one can, at least in principle,
transmit data in a totally secure fashion \cite{eke91},
overcome (double) the limit of classical information that can
be transmitted from one region to another by sending the same
number of particles \cite{ben92} and, not least
fantastic, teleport microscopic particles \cite{ben93,bra98}.

These and many other quantum informational tasks \cite{livrodozeilinger}
assume the existence of perfect entanglement among 
the many particles that constitute a given quantum system.
In other words, one needs maximally entangled states for the successful 
implementation of several of the previous protocols. However,
perfect entanglement is extremely difficult to generate 
and preserve in the laboratory, particularly when one 
deals with many particles at the same time; 
maximally entangled states rapidly degrade to partially entangled ones.  

A possible solution to this problem is to develop entanglement distillation 
strategies \cite{ben96}. By entanglement distillation
we mean any protocol whose ultimate goal is to obtain a 
maximally entangled states from a given number of partially
entangled ones. In general, a distillation procedure is only useful if 
it operates exclusively through local operations and classical communication
(LOCC). Local operations are, for example, single measurements on subsystems or
local unitary transformations. For classical communication
we mean that there is the possibility of exchanging classical signals between
subsystems using classical communication channels. For example, 
the result of a measurement at a given location can be transmitted by radio waves
to the sites where the other subsystems lie.

However, we may ask whether there could be another
way to explore partially entangled states in order to make them useful
to the implementation of the aforementioned protocols. Putting it another
way, isn't it
possible to use them \textit{directly} to perform an informational
task, without relying on distillation protocols? Moreover, 
it might happen that we only have at our disposal one copy of
a partially entangled state and do not have the resources to implement a distillation
protocol. Is there not a way to directly use it
to accomplish our task?
For some communication protocols, the answers to the previous questions
are positive if one deals with pure partially entangled states 
\cite{Guo00,Agr02,Gor06A,Gor06B,Gor07,Gor10,Mod08,Rig09}. 

In this paper we further explore the direct use of partially entangled 
states to perform quantum communication tasks. First we review the 
three major strategies to harness partially entangled states for direct quantum
teleportation. We then compare their efficiencies assuming either perfect or noisy devices
(non-perfect measurements and unitary evolutions). By doing
so we give conditions under which a particular strategy is better suited than
the other two. Second, we present new quantum teleportation protocols which are built
combining those three major 
techniques. This leads to an increase in the efficiency of those well known protocols 
\cite{Guo00,Agr02,Gor06A,Gor06B,Gor07,Gor10,Mod08,Rig09}.  
We also present a direct multiple
teleportation protocol based on partially entangled states  
whose efficiency is exactly equal to the best known distillation protocol 
devised to accomplish the same task \cite{Per08}. 
We then develop new teleportation protocols employing multipartite partially
entangled states as our quantum resource \cite{Agr06}, in contrast to standard approaches
that use partially entangled Bell states. The three techniques are also 
combined here in order to achieve the highest efficiency possible.
We then show that among the new protocols there exist three cases where
the teleportation is successful without knowing the entanglement content 
of the partially entangled quantum channels, as is mandatory in Refs. \cite{Guo00,Agr02} 
and in some cases of Ref. \cite{Rig09}. Finally, we show how one can 
compute the efficiency upper bound for direct teleportation protocols that make
use of partially entangled Bell states.

\section{The standard techniques}
\label{sec2}

As anticipated in the introductory paragraphs,
maximally entangled states are not really
essential for the implementation of many 
informational tasks. However, there is a price to pay. Although
the cost to produce and preserve maximally entangled states is high, 
any protocol that makes use of these states
theoretically work with probability one. On the other hand, by replacing maximally
entangled states with partially entangled ones, one faces, even theoretically,
instances in which the protocol fails.
In other words, when using maximally entangled states one deals
with deterministic protocols, while using partially entangled states
one necessary has probabilistic ones.

Quantum communication protocols, and in particular teleportation,
directly using partially entangled states
can be divided into at least
three groups. In each group one uses a 
fundamentally different strategy to cope with
the limitation of dealing
with partially entangled states.
In the first group \cite{Guo00} the imperfection caused
by using a non-ideal channel is corrected by interacting
an auxiliary qubit with the qubit carrying the final imperfect
teleported state. By correctly choosing the interaction
between them, which depends on
the  partially entangled state being used in the protocol as
well as on the measurement results of Alice, one can transfer 
the error to the auxiliary qubit and 
``clean'' the qubit carrying the teleported state.

The second group includes strategies that modify
the original teleportation protocol \cite{ben93}. In particular, one changes
the measurements made onto the input qubits of the protocol
\cite{Agr02,Gor06A,Gor06B,Gor10}. Depending on the type of
partially entangled state used, specific changes
to the protocol turns it useful again.

Finally, we have the third group \cite{Mod08,Rig09}, which contains the
cases where one repeats the same protocol, one or more times, using
as input the output of the previous implementation.
In this way one gets a correct result after a few repetitions.
In this group one has the phenomenon of self-correction, where
an error introduced in one step is compensated by the next error
in the next step of the protocol.

So far, however, little has been done in the direction to combine
these three strategies to increase the success rate of teleportation
protocols based on the direct use of partially entangled states. 
A study in which two of these techniques
were combined with success as well as the introduction of another one was made in
Ref. \cite {Rig09}. One of our goals here is to combine these many strategies in order
to improve the efficiency of the most well-known probabilistic teleportation
protocols \cite{Guo00,Agr02,Gor06A,Mod08,Rig09}.

\begin{figure}[!ht]
\begin{center}
\includegraphics[width=8cm]{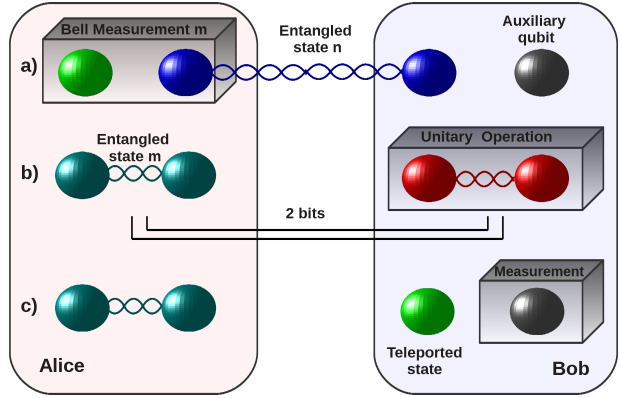}
\end{center}
\caption{\label{fig1}  a) Alice and Bob share the entangled
state $|\Phi_n^+\rangle$ and she performs a generalized Bell measurement 
spanned by the states labeled by $m$; b) Alice informs Bob of her measurement
result. This information allows Bob to properly choose what sort of unitary 
operation (interaction) $\mathbf{U}$ 
to implement onto the recipient and ancilla qubits and also which 
local unitaries (Pauli matrices) to apply onto the recipient qubit; 
c) Bob measures the ancilla.
Depending on his result, the recipient qubit is 
exactly given by the state that described Alice's original qubit.
In the original protocol \cite{ben93} $m=n=1$ and there is no ancilla. For 
group $1$ \cite{Guo00} $m=1$, $n<1$, and $\mathbf{U}=\mathbf{U}_n$ while for 
group $2$ \cite{Agr02} $m=n<1$ and no ancilla is used (or equivalently $\mathbf{U}=\mathbb{1}$). 
For group $3$ \cite{Mod08,Rig09} there is no ancilla and
the teleportation protocol with $n<1$ and $m=1$ is implemented twice.}
\end{figure}

In order to be self contained and to pave the way to the presentation of our
new results in the next sections, 
we now present in a unified formalism the three strategies just mentioned.
See Fig. \ref{fig1} for a schematic unified view of all standard probabilistic
teleportation protocols.

Let the qubit Alice wants to teleport to Bob be described by
\begin{equation}
|\phi\rangle_1=\alpha|0\rangle_1 + \beta|1\rangle_1,
\label{eq:qubitoriginaldenovo}
\end{equation}
with the normalization condition $|\alpha|^2+|\beta|^2=1$,
and the entanglement resource shared between Alice and Bob 
(quantum channel) be given by
\begin{equation}
|\Phi^+_n\rangle_{2,3}=\frac{|00\rangle_{2,3}
+n|11\rangle_{2,3}}{\sqrt{1+n^2}},
\label{canaln}
\end{equation}
with $0\leq n\leq1$. When $n=0$ there is no entanglement while
for $n=1$ we obtain a maximally entangled state. The subscripts
label the qubits, where qubits $1$ and $2$ are with Alice and $3$
with Bob.

We also define the generalized Bell states ($B_m$) as \cite{Guo00,Agr02,Gor06A} 
\begin{eqnarray}
|\Phi^+_m\rangle=\frac{|00\rangle + m|11\rangle}{\sqrt{1+m^2}}, 
|\Phi^-_m\rangle=\frac{m|00\rangle -|11\rangle}{\sqrt{1+m^2}},
|\Psi^+_m\rangle=\frac{|01\rangle +m|10\rangle}{\sqrt{1+m^2}}, 
|\Psi^-_m\rangle=\frac{m|01\rangle - |10\rangle}{\sqrt{1+m^2}},
\label{basegeneralizada}
\end{eqnarray}
where $0\leq m\leq1$. 

Inverting Eq.~(\ref{basegeneralizada}) we can express the standard
basis $\{|00\rangle, |01\rangle, |10\rangle, |11\rangle\}$ as functions
of $B_m$. This implies that the three qubit state can be written as
%
%
\begin{eqnarray}
|\Psi\rangle_{123}&=&|\phi\rangle_{1}|\Phi^+_n\rangle_{2,3}
=\eta_1(\alpha,\beta)|\Phi^+_m\rangle_{1,2}
|\phi_1\rangle_3
+\eta_2(\alpha,\beta)|\Phi^-_m\rangle_{1,2}
|\phi_2\rangle_3
\nonumber\\
&&+\eta_2(\beta,\alpha)|\Psi^+_m\rangle_{1,2} 
|\phi_3\rangle_3
+\eta_1(\beta,\alpha)|\Psi^-_m\rangle_{1,2} 
|\phi_4\rangle_3,\nonumber \\
\label{eq:estgeralindiano1}
\end{eqnarray}
where  
\begin{eqnarray}
\eta_1(\alpha,\beta)=\sqrt{\frac{|\alpha|^2+m^2n^2|\beta|^2}
{(1+m^2)(1+n^2)}},\,\,\,
\eta_2(\alpha,\beta)=\sqrt{\frac{m^2 |\alpha|^2+n^2|\beta|^2}
{(1+m^2)(1+n^2)}},
\end{eqnarray}
and the normalized states with Bob are
\begin{eqnarray}
|\phi_1\rangle_3=
\frac{\alpha|0\rangle_3 + mn\beta|1\rangle_3}
{\sqrt{|\alpha|^2+m^2 n^2 |\beta|^2}},
\,\,\,
|\phi_2\rangle_3=
\frac{m \alpha|0\rangle_3 - n\beta|1\rangle_3}
{\sqrt{m^2 |\alpha|^2+n^2 |\beta|^2}},
\\
|\phi_3\rangle_3=
\frac{m \beta|0\rangle_3 + n\alpha|1\rangle_3}
{\sqrt{n^2 |\alpha|^2+m^2 |\beta|^2}},
\,\,\,
|\phi_4\rangle_3=
\frac{-\beta|0\rangle_3 + m n \alpha|1\rangle_3}
{\sqrt{m^2 n^2 |\alpha|^2+|\beta|^2}}.
\end{eqnarray}

So far no physical processes were implemented. We have just
rewritten the global state describing Alice's and Bob's qubits.
The generalized teleportation protocol begins by Alice 
implementing a joint measurement onto her two qubits in the 
$B_m$ basis.
The probability to get a particular $B_m$ and the
respective collapsed 
state after the measurement is given by
\begin{eqnarray}
P_{|\Phi^+_m\rangle} = \eta_1(\alpha,\beta)^2 \longrightarrow 
|\Phi^+_m\rangle_{1,2}|\phi_1\rangle_3, \,\,\,
P_{|\Phi^-_m\rangle} = \eta_2(\alpha,\beta)^2 \longrightarrow
|\Phi^-_m\rangle_{1,2}|\phi_2\rangle_3, \label{p2}\\
P_{|\Psi^+_m\rangle} = \eta_2(\beta,\alpha)^2 \longrightarrow
|\Psi^+_m\rangle_{1,2} |\phi_3\rangle_3, \,\,\,	
P_{|\Psi^-_m\rangle} = \eta_1(\beta,\alpha)^2 \longrightarrow
|\Psi^-_m\rangle_{1,2} |\phi_4\rangle_3.\label{p4}
\end{eqnarray}

Alice then informs Bob about her measurement result which allows
him to implement the corresponding unitary operations (Pauli matrices)
as listed in the Tab. \ref{tab:corrpauli}.
\begin{table}[!htb]
\caption{\label{tab:corrpauli}
Unitary operations Bob must implement on his qubit after
being informed of Alice's measurement result.}
\begin{center}
\begin{tabular}{cc}
Alice's measurement result  & Bob's correction \\ \hline 
$|\Phi^+_m\rangle$ & $\mathbf{1}$ \\
$|\Phi^-_m\rangle$ & $\sigma_{z}$ \\
$|\Psi^+_m\rangle$ & $\sigma_{x}$ \\
$|\Psi^-_m\rangle$ & $\sigma_{z}\sigma_{x}$ \\
\hline
\end{tabular}
\end{center}
\end{table}   

After Alice's measurement and after Bob implements the Pauli 
unitary operations his qubit is described by one of the following
four possibilities,
\begin{eqnarray}
|\phi_1\rangle_3\longrightarrow
|\varphi_1\rangle_3=\frac{\alpha|0\rangle_3 + mn\beta|1\rangle_3}
{\sqrt{|\alpha|^2+m^2 n^2 |\beta|^2}},
\,\,\,
|\phi_2\rangle_3\longrightarrow
|\varphi_2\rangle_3=\frac{m \alpha|0\rangle_3 + n\beta|1\rangle_3}
{\sqrt{m^2 |\alpha|^2+n^2 |\beta|^2}},
\label{phi2}
\\
|\phi_3\rangle_3\longrightarrow
|\varphi_3\rangle_3=\frac{n\alpha|0\rangle_3 + m \beta|1\rangle_3}
{\sqrt{n^2 |\alpha|^2+m^2 |\beta|^2}},
\,\,\,
|\phi_4\rangle_3\longrightarrow
|\varphi_4\rangle_3=\frac{m n \alpha|0\rangle_3 + \beta|1\rangle_3}
{\sqrt{m^2 n^2 |\alpha|^2+|\beta|^2}}.
\label{phi4}
\end{eqnarray}

Note that when $m=n=1$ we recover the original teleportation
protocol \cite{ben93}. Indeed, for such conditions 
$P_{|\Phi^{\pm}_m\rangle}=P_{|\Psi^{\pm}_m\rangle}=1/4$ and 
$|\phi_j\rangle_3=|\phi\rangle$, $j=1,\ldots, 4$, giving 
a total probability of success ($P_{suc}$) equals to one.

\subsection{Group 1: interaction with an ancilla}

If we set $m=1$ with $n<1$ we recover group $1$ \cite{Guo00}. With such a choice
the teleported state is never identical to Alice's qubit $1$. 
Looking at Eqs.~(\ref{phi2})-(\ref{phi4}) we see that 
Bob's final state (unnormalized) 
is either $\alpha|0\rangle+n\beta|1\rangle$ or $n\alpha|0\rangle+\beta|1\rangle$.
In order to get rid of the $n$ multiplying either $\alpha$ or $\beta$ we interact
this qubit with an auxiliary one (an ancilla) initially 
described by $|0\rangle_{aux}$.  For states
given by $\alpha|0\rangle+n\beta|1\rangle$ the unitary matrix (in the standard basis)
describing the interaction is
\begin{equation}
\mathbf{U}_{n} = 
\left(
\begin{array}{cccc}
n & \sqrt{1-n^2} & 0 & 0\\
0 & 0 & 0 & 1\\
0 & 0 & 1 & 0\\
\sqrt{1-n^2} & -n & 0 & 0 
\end{array} 
\right).
\label{matrizunit}
\end{equation}

Thus, assuming Alice measures $|\Phi^+_m\rangle$, Bob's qubits after the interaction become 
\begin{eqnarray}
\mathbf{U}_n|\varphi_1\rangle_3|0\rangle_{aux}&=&
\frac{n}{\sqrt{|\alpha|^2+n^2|\beta|^2}}
(\alpha|0\rangle_3+\beta|1\rangle_3)|0\rangle_{aux}
+\sqrt{\frac{1-n^2}{|\alpha|^2+n^2|\beta|^2}}\alpha|1\rangle_3|1\rangle_{aux}.
\label{corrfinalchines}
\end{eqnarray} 
We now see that if Bob measures the ancilla in the standard basis and he gets $|0\rangle_{aux}$
the other qubit collapses to the correct state and the teleportation is successful.
If he gets $|1\rangle_{aux}$ the protocol fails and no information about Alice's original
qubit is retained. A similar analysis applies for $|\varphi_2\rangle_3$ (Bob's state if
Alice measures $|\Phi^-_m\rangle$). 

In those cases the probability for a single successful event is given by 
\begin{equation}
P_U=P_{|\Phi^+_m\rangle}P_{|0\rangle_{aux}} = P_{|\Phi^-_m\rangle}P_{|0\rangle_{aux}}=
\frac{n^2}{2(1+n^2)},
\end{equation}
where $P_{|0\rangle_{aux}}=n^2/(|\alpha|^2+n^2|\beta|^2)$ is the probability
of measuring $|0\rangle_{aux}$. 

For the other set of states given by $n\alpha|0\rangle+\beta|1\rangle$ 
the unitary matrix describing the interaction is
\begin{equation}
\mathbf{V}_{n} = 
\left(
\begin{array}{cccc}
1 & 0 & 0 & 0\\
0 & 0 & 0 & 1\\
0 & \sqrt{1-n^2} & n & 0\\
0 & -n & \sqrt{1-n^2} & 0 
\end{array} 
\right).
\label{matrizunit2}
\end{equation}

Bob's qubits after the interaction become (assuming Alice measures $|\Psi^+_m\rangle$) 
\begin{eqnarray}
\mathbf{V}_n|\varphi_3\rangle_3|0\rangle_{aux}=
\frac{n}{\sqrt{n^2|\alpha|^2+|\beta|^2}}
(\alpha|0\rangle_3+\beta|1\rangle_3)|0\rangle_{aux}
+\sqrt{\frac{1-n^2}{n^2|\alpha|^2+|\beta|^2}}\beta|1\rangle_3|1\rangle_{aux}.
\label{corrfinalchines2}
\end{eqnarray} 
If Bob measures the ancilla and obtains $|0\rangle_{aux}$
the other qubit collapses to the correct state and the teleportation is successful.
If he gets $|1\rangle_{aux}$ the protocol fails. 
A similar analysis applies for state $|\varphi_4\rangle_3$.

In those two cases the probability for a single successful event is given by 
\begin{equation}
P_V=P_{|\Psi^+_m\rangle}\tilde{P}_{|0\rangle_{aux}} = P_{|\Psi^-_m\rangle}\tilde{P}_{|0\rangle_{aux}}=
\frac{n^2}{2(1+n^2)},
\end{equation}
where $\tilde{P}_{|0\rangle_{aux}}=n^2/(n^2|\alpha|^2+|\beta|^2)$ is the probability
of measuring $|0\rangle_{aux}$. 

Finally, for a single run of the protocol 
the total probability of success is
\begin{equation}
P_{suc_1}= 2P_U+2P_V=\frac{2n^2}{1+n^2},
\label{group1}
\end{equation}
where the subindex $1$ reminds us that this value is associated to the first
group of strategies to directly use partially entangled states in the execution of
a quantum informational task.

\subsection{Group 2: generalized Bell measurements}

If Alice chooses $m=n$ (matching condition), i.e., she implements
a generalized Bell measurement, we get
a probabilistic teleportation protocol belonging
to group $2$ \cite{Agr02}. In this case, looking at Eqs.~(\ref{p2})-(\ref{p4}) 
and (\ref{phi2})-(\ref{phi4}), we realize that whenever Alice measurements yield 
$|\Phi_m^-\rangle$ and $|\Psi^+_m\rangle$ the protocol is successful and Bob's state
is exactly described by $|\phi\rangle$ (Eq.~(\ref{eq:qubitoriginaldenovo})).
In this case, 
\begin{equation}
P_{suc_2}=P_{|\Phi^{-}_m\rangle}+P_{|\Psi^{+}_m\rangle}=\frac{2n^2}{(1+n^2)^2},
\label{group2}
\end{equation}
where the subindex $2$ reminds us that this value is associated to the second
group. Note that $P_{suc_1}>P_{suc_2}$.

\subsection{Group 3: repeated teleportation}

The multiple teleportation protocols of Ref. \cite{Mod08,Rig09} is obtained
by implementing twice the teleportation protocol with $m=1$ and $n<1$. And
we can understand how it works as follows. 
For a detailed analysis, with all possible output states and different measurement
basis and channels, the reader
is directed to \ref{appendixA} and to Ref. \cite{Rig09}.

If we look at
Eqs.~(\ref{p2})-(\ref{phi4}) we note that whenever Alice measures $|\Phi^{\pm}_m\rangle$
the final state (unnormalized) with Bob is $\alpha_j|0\rangle+n\beta_j|1\rangle$ while
if Alice measures $|\Psi^{\pm}_m\rangle$ Bob's state goes to $n\alpha_j|0\rangle+\beta_j|1\rangle$.
Here $\alpha_j$ and $\beta_j$ are the coefficients of Bob's state before the $j$-th teleportation.
Therefore, if after the first teleportation Alice's measurement yielded $|\Phi^{\pm}_m\rangle$ 
($|\Psi^{\pm}_m\rangle$), the second teleportation will be successful 
if she measures $|\Psi^{\pm}_m\rangle$ ($|\Phi^{\pm}_m\rangle$). Indeed, for any of 
these eight possibilities, which occur with equal chances,
the final state (unnormalized) with Bob will be given by $n(\alpha|0\rangle+\beta|1\rangle)$.
This leads to the following overall probability of success,
\begin{equation}
P_{suc_3}=\frac{2n^2}{(1+n^2)^2}.
\label{group3}
\end{equation}

Note that $P_{suc_3}=P_{suc_2}<P_{suc_1}$. Also, this strategy assumes that the second entangled
state has the same degree of entanglement of the first one and that it belongs to Bob.

\subsection{Comparison among the three groups assuming imperfect devices}
\label{sec2d}

So far all previous works assumed that the Bell measurements
and the unitary corrections $\mathbf{U}_n$ and $\mathbf{V}_n$ are free from imperfections.
In this case it is simple to see that the strategy $1$ gives the best result. 
Let us introduce now imperfect measurements and imperfect unitary corrections in order to compare
the efficiencies of all three strategies. Whenever a failure is noticed, the photon detector does
not click, for example, we have to discard a possible theoretically successful run of the
protocol. Therefore, we will see a decrease in the overall probability of success.  

In a simple but realistic situation, the decrease of the 
probability of success for each one of the three groups can be modeled as follows,
\begin{eqnarray}
P_{suc_1}=\epsilon_b\epsilon_u\frac{2n^2}{(1+n^2)},\,\,\,
P_{suc_2}=\epsilon_m\frac{2n^2}{(1+n^2)^2},\,\,\,
P_{suc_3}=\epsilon_b^2\frac{2n^2}{(1+n^2)^2}.
\label{group123}
\end{eqnarray}
Here $0\leq\epsilon_j\leq1$, $j=b,u,m$. $\epsilon_b$ is the reduction in efficiency due to an imperfect Bell measurement,
$\epsilon_u$ is the reduction due to an imperfect unitary correction as given in group $1$, 
and $\epsilon_m$ is the reduction in efficiency due to an imperfect generalized Bell measurement.

\begin{figure}[!ht]
\begin{center}
\includegraphics[width=7cm]{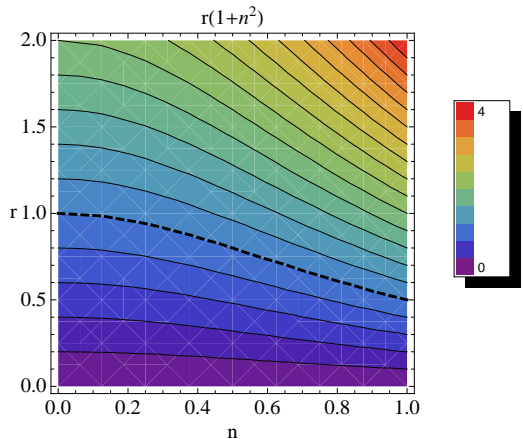}
\end{center}
\caption{\label{fig1a}  $P_{suc_1}/P_{suc_j}$, $j=2,3$, as a function of $n$ and $r$. $n$ is
related to the degree of entanglement of the channel and $r$ to the errors in a realistic implementation of
the protocols. See text for details. The dashed line represents the threshold ($P_{suc_1}/P_{suc_j}=1$)
above which $P_{suc_1}>P_{suc_j}$. In other words, it delimits the region
above which the strategies of group $1$ outperform the others.}
\end{figure}

The first group continues to be the best strategy for a given $n$ whenever $P_{suc_1}/P_{suc_2}>1$ and
$P_{suc_1}/P_{suc_3}>1$ which leads respectively to $\epsilon_b\epsilon_u(1+n^2)/\epsilon_m>1$ and
$\epsilon_u(1+n^2)/\epsilon_b>1$. Both conditions are of the form $r(1+n^2)>1$, with $r$ being respectively
$\epsilon_b\epsilon_u/\epsilon_m$ and $\epsilon_u/\epsilon_b$. The values of $r$ and $n$ 
in which $P_{suc_1}$ is better than
both strategies are depicted in Fig. \ref{fig1a}. 
To complete the analysis, $P_{suc_2}/P_{suc_3}=\epsilon_m/\epsilon_b^2>1$ is the condition under
which the second strategy outperforms the third one in a realistic setting.

\section{Combining the three standard techniques}

To the best of our knowledge the first attempt to combine those techniques in a systematic manner
was done in Ref. \cite{Rig09}. However, in the protocols presented there
only the last two of the three aforementioned strategies (group $2$ and group $3$)
were simultaneously employed. Our goal here is to investigate if and how the use of the 
first strategy can lead to an improvement in the performance of the most successful 
protocol of Ref. \cite{Rig09}.

\begin{figure}[!ht]
\begin{center}
\includegraphics[width=7cm]{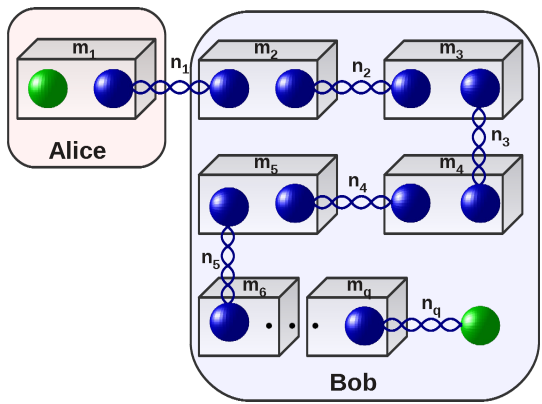}
\end{center}
\caption{\label{fig2}  Here Alice and Bob share only one entangled state
labeled by $n_1=n$ (Eq.~(\ref{canaln})). Boxes denote Bell measurements. The three protocols in Ref. \cite{Rig09} that
assume this configuration can be described as follows. Protocol $1$: Alice and Bob
realize standard Bell measurements ($m_j=1$, all $j$) and all quantum channels possess the
same entanglement ($n_j=n$, all $j$). In all protocols Bob stops when he gets a successful event, i.e.,
when after the $q$-th teleportation his state is $\alpha|0\rangle + \beta|1\rangle$.
Protocol $2$: The quantum channels are the same as before but now Alice and Bob realize
generalized Bell measurements ($m_j=n$, all $j$). Protocol $3$: Now Alice and Bob get back
to standard Bell measurements ($m_j=1$, all $j$) but the entangled states have their 
entanglement reduced according to the following rule: $n_1=n$, $n_2=n$, $n_3=n^2$, $n_4=n^4$,
$\ldots$, $n_q=n^{2^{q-2}}$. Out of these three protocols, this last one is the most efficient.}
\end{figure}

Before we proceed it is instructive to review the three protocols given in \cite{Rig09}, with
special care to the third and most efficient one. For further details the reader is directed
to Ref. \cite{Rig09}. In the protocols we will be dealing with in this section
it is assumed that Alice and Bob share only one partially entangled state as given by 
Eq.~(\ref{canaln}). All other partially entangled
states are with Bob, who may use them or not to implement subsequent 
teleportations to himself, depending on which strategy he chooses to employ to get rid of the error
introduced from the first teleportation.  See Fig. \ref{fig2} for more details. 

The first protocol, which is equivalent to the one of Ref. \cite{Mod08}, is just 
successive applications of the strategy described in group $3$. Bob keeps teleporting
the outcome from the first teleportation until he gets a sequence of Bell
measurements that makes the output of the $q$-th teleportation equals to 
$\alpha|0\rangle+\beta|1\rangle$, the original state with Alice. We can
understand it looking at 
Tab. \ref{apptable1} in \ref{appendixA}.
For $m_{j}=1$ and $n_{j}=n$ we readily see that 
after the $j$-th teleportation and whenever
the Bell measurement outcome is $|\Phi_{m_{j}}^{\pm}\rangle$ the
coefficients of the input state (the qubit before the $j$-th teleportation)
change as ($\alpha_{j}
\rightarrow  \alpha_{j}$, $\beta_{j}\rightarrow n \beta_{j}$).
On the other hand, whenever the Bell measurement results in 
$|\Psi_{m_{j}}^{\pm}\rangle$ we have 
($\alpha_{j} \rightarrow n \alpha_{j}$, $\beta_{j}\rightarrow
\beta_{j}$). Therefore, we always get $\alpha|0\rangle + \beta|1\rangle$
when a balanced number of $|\Phi^{\pm}_{m_j}\rangle$ and $|\Psi^{\pm}_{m_j}\rangle$ appears
in a sequence of measurements.  

The second protocol of \cite{Rig09} combines the strategies of groups $2$ and $3$.
Here we have all teleportations with channels given by $n_j=n$ and generalized
Bell measurements with the matching condition ($m_j=n_j$). In this scenario,
Tab. \ref{apptable1} tells us that (a) the coefficients of Bob's qubit 
after the $j$-th teleportation is modified according to the rule 
$(\alpha_{j},\beta_j) \rightarrow (\alpha_{j}, n^2\beta_{j})$
whenever $\ket{\Phi^+_{m_j}}$ is a result of the generalized Bell
measurement; (b) $(\alpha_{j},\beta_j) \rightarrow n(\alpha_{j},\beta_{j})$
if the measurement result is $\ket{\Phi^-_{m_j}}$ or $\ket{\Psi^+_{m_j}}$; and
(c) $(\alpha_{j},\beta_j) \rightarrow (n^2\alpha_{j},\beta_{j})$
if the outcome is $\ket{\Psi^-_m}$.
Hence, for an equal number of $|\Phi^{+}_{m_j}\rangle$
and $\ket{\Psi^{-}_{m_j}}$, $m_j=n_j$, in a sequence of generalized
Bell measurements Bob gets $\alpha|0\rangle + \beta|1\rangle$. 
The $n^2\beta_j$ coming from the measurement of
$|\Phi^{+}_{m_j}\rangle$ is balanced by the $n^2\alpha_j$ coming
from another measurement giving $\ket{\Psi^-_{m_j}}$. The states
$\ket{\Phi^-_{m_j}}$ and $\ket{\Psi^+_{m_j}}$ are `neutral', multiplying
both $\alpha_j$ and $\beta_j$ by $n$.
As was shown in \cite{Rig09} this second protocol achieves the same
efficiency of the first one employing, nevertheless, just \textit{half} 
of the number of teleportations. This economy on entanglement resources
is important on its on sake and also for practical implementations 
since it reduces other possible errors introduced by imperfect projective
measurements as described in Sec. \ref{sec2d}. 

The third protocol is based on standard Bell
measurements ($m_j=1$, all $j$) 
and, differently from the two previous protocols, at each teleportation 
the entanglement of the quantum channel
is \textit{changed} according to the following rule: $n_{j}=n_{j-1}^2$,
$j\geq 3$, while $n_1=n_2=n$. In other words, 
after the first two teleportations we \textit{decrease} the entanglement of 
the quantum channel at each subsequent teleportation. The first two steps of this protocol 
are exactly the same of protocol $1$.  
After the second teleportation,
the unsuccessful instances are described by the (unnormalized) state
$\alpha\ket{0}+n^2\beta\ket{1}$, if the two Bell measurements yield 
$|\Phi^{\pm}_{m_1}\rangle|\Phi^{\pm}_{m_2}\rangle$, or by 
$n^2\alpha\ket{0}+\beta\ket{1}$, if we have
$|\Psi^{\pm}_{m_1}\rangle|\Psi^{\pm}_{m_2}\rangle$. 
In order to catch up with the $n^2$ multiplying either $\alpha$
or $\beta$ in the next run of the protocol we use the quantum
channel $\ket{\Phi^+_{n_3}}$, with $n_3=n^2$.
In this case the previous teleported qubit changes to 
$(\alpha_3,\beta_3) \rightarrow (\alpha_{3}, n^2\beta_{3})$ if we measure
$|\Phi^{\pm}_{m_3}\rangle$ or to $(\alpha_{3},\beta_3) \rightarrow
(n^2\alpha_{3}, \beta_{3})$ if we get $|\Psi^{\pm}_{m_3}\rangle$. 
Note that $(\alpha_j,\beta_j)$ are the input state for the $j$-th
run of the protocol (the qubit before the $j$-th teleportation). Thus,
whenever the following sequences of Bell measurements occur,
$|\Phi^{\pm}_{m_1}\rangle|\Phi^{\pm}_{m_2}\rangle|\Psi^{\pm}_{m_3}\rangle$ or
$|\Psi^{\pm}_{m_1}\rangle|\Psi^{\pm}_{m_2}\rangle|\Phi^{\pm}_{m_3}\rangle$, 
Bob's output are exactly $\alpha|0\rangle+\beta|1\rangle$.

This same reasoning can be repeated at each new teleportation. 
It is not difficult to see that after the $(q-1)$-th teleportation the
unsuccessful instances are given by one of the following 
$2\times 2^{q-1}$ sequences of Bell measurements: 
$\otimes_{j=1}^{q-1}\ket{\Phi^{\pm}_{m_j}}$ or
$\otimes_{j=1}^{q-1}\ket{\Psi^{\pm}_{m_j}}$. For sequences involving 
$\ket{\Phi^{\pm}_{m_j}}$ the (unnormalized) teleported
qubit is described by $\alpha \ket{0} + n^{2^{q-2}}\beta\ket{1}$. For the second
set of sequences by $n^{2^{q-2}} \alpha \ket{0} + \beta\ket{1}$. 
After the $q$-th teleportation, though, we are successful if the sequence
of Bell measurements belongs to either one of the following sets, 
$(\otimes_{j=1}^{q-1}\ket{\Phi^{\pm}_{m_j}})\ket{\Psi^{\pm}_{m_q}}$ or
$(\otimes_{j=1}^{q-1}\ket{\Psi^{\pm}_{m_j}})\ket{\Phi^{\pm}_{m_q}}$, where 
we have employed the quantum channel $\ket{\Phi^{+}_{n_{q}}}$, with $n_{q}=n^{2^{q-2}}$.

The probability to get any one of those sequences are identical and if we
take into account that we have $2 \times 2^q$ possible successful sequences at 
the $q$-th teleportation we
obtain \cite{Rig09} 
\begin{equation}
P_{suc}^{(q)}=\frac{2(1-n^2)n^{2^{q-1}}}{(1+n^2)(1-n^{2^{\,\!^q}})}
\end{equation}
for the probability of success at the $q$-th teleportation ($q\geq2$). 
The total probability of success after $q$ teleportations is therefore
\begin{equation}
P_{suc}^{tot}(n,q)=\sum_{j=1}^q P_{suc}^{(j)},
\label{total1}
\end{equation}
where we define $P_{suc}^{(1)}=0$ and make it explicit that $P_{suc}^{tot}$ depends
on the entanglement of the initial channel ($n$) and on the
number of teleportations ($q$).
Note that this protocol also introduces a new strategy that does not
belong to groups $1$, $2$ or $3$ of Sec. \ref{sec2}. 
Indeed, this
third protocol uses entangled channels with less entanglement 
at each new teleportation. And in \cite{Rig09} it was shown that 
among the three protocols just described, and for a fixed number of teleportations $q$, 
this protocol achieves the highest efficiency (greater $P_{suc}^{tot}(n,q)$) 
spending the least amount of entanglement. 

Our goal, now, is to show that we can improve even more the efficiency of this last protocol by
applying the strategies described in group $1$, namely, the interaction of the
output qubit after the $q$-th teleportation with an ancilla. We also tested how its efficiency
would change if we applied the strategies of group $2$ (generalized Bell measurements) 
at each teleportation, imposing the matching condition ($m_j=n_j$). This approach, however, 
led to no improvement of the efficiency of protocol $3$.

After the $q$-th teleportation the probability of success for this protocol
is given by Eq.~(\ref{total1}). Instead of continuing with another teleportation
using the quantum channel $\ket{\Phi^{+}_{n_{q+1}}}$, with $n_{q+1}=n^{2^{q-1}}$, we
apply the techniques of group $1$ to correct those cases that failed. The cases of failure
can be divided into two sets, according to the following sequences of Bell measurements, 
$\otimes_{j=1}^{q}\ket{\Phi^{\pm}_{m_j}}$ and $\otimes_{j=1}^{q}\ket{\Psi^{\pm}_{m_j}}$.
Hence, there are $2^q$ cases for each set giving a total of $2\times 2^q$ possible
sequences of Bell measurements that lead to a failure. 

The unsuccessful states with Bob after the $q$-th teleportation are given by 
$(\alpha|0\rangle+n^{2^{q-1}}\beta|1\rangle)/$ $\sqrt{|\alpha|^2+n^{2^{q}}|\beta|^2}$
if the sequence of Bell measurements belongs to the set $\otimes_{j=1}^{q}\ket{\Phi^{\pm}_{m_j}}$
or by $(n^{2^{q-1}}\alpha|0\rangle+\beta|1\rangle)/\sqrt{n^{2^{q}}|\alpha|^2+|\beta|^2}$ if 
we have sequences belonging to the other set. From now on we restrict our analysis to the first
set of Bell measurements since all that we obtain below applies to the other set 
by simply exchanging $\alpha \leftrightarrow \beta$, with the final result equal to
the one of the first set and independent of $\alpha$ and $\beta$.

The probability to obtain a given sequence of the first set are all equal and given by
the product of the probability to get at each teleportation $j$ either state 
$\ket{\Phi^{+}_{m_j}}$ or $\ket{\Phi^{-}_{m_j}}$ as the outcome of a Bell measurement. 
These probabilities for the first teleportation
are given by Eq.~(\ref{p2}) and they are equal since $m_j=1$, any $j$. 
The calculations that led to (\ref{p2}) are very similar to the ones needed
at each new teleportation. Only two things change. First, before the $j$-th teleportation
the input state is 
$(\alpha|0\rangle+n^{2^{j-2}}\beta|1\rangle)/\sqrt{|\alpha|^2+n^{2^{j-1}}|\beta|^2}$
instead of $(\alpha|0\rangle+\beta|1\rangle)$. Second, the quantum channel at each 
subsequent teleportation changes according to the rule previously given. Taking these two
facts into account we can show that 
\begin{equation}
P_{|\Phi^{\pm}_{m_j}\rangle} = \frac{|\alpha|^2+n^{2^{j}}|\beta|^2}
{2(1+n^{2^{j-1}})(|\alpha|^2+n^{2^{j-1}}|\beta|^2)}, \hspace{.05cm} \mbox{for} \hspace{.05cm}
j\geq 2,\label{ProbA}
\end{equation}
is the probability to measure $|\Phi^{\pm}_{m_j}\rangle$ at the $j$-th teleportation.

Also, since after the $q$-th teleportation the state with Bob is 
$(\alpha|0\rangle+n^{2^{q-1}}\beta|1\rangle)/\sqrt{|\alpha|^2+n^{2^{q}}|\beta|^2}$
we can correct it using the strategies of group $1$. We interact it with an ancilla
$|0\rangle_{aux}$ via the unitary operation given by Eq.~(\ref{matrizunit}) with
$n\rightarrow n^{2^{q-1}}$. This leads to the following probability of
``cleaning'' the state to $\alpha|0\rangle + \beta|1\rangle$,
\begin{equation}
P^{(q+1)}_{|0\rangle_{aux}}=\frac{n^{2^{q}}}{|\alpha|^2+n^{2^{q}}|\beta|^2}.
\end{equation}

Thus, if we take into account all $2\times 2^q$ cases we obtain
\begin{equation}
\tilde{P}^{(q+1)}_{suc}=2^{q+1}\prod_{j=1}^{q}P_{|\Phi^{\pm}_{m_j}\rangle}
P^{(q+1)}_{|0\rangle_{aux}}=\frac{2(1-n^2)n^{2^{q}}}{(1+n^2)(1-n^{2^{q}})}
\label{total2}
\end{equation}
as the probability of getting a clean state after implementing the
strategies of group $1$ to the unsuccessful cases after the $q$-th teleportation. 
Here $P_{|\Phi^{\pm}_{m_1}\rangle}$ is given
by Eq.~(\ref{p2}) with $m=1$ and $P_{|\Phi^{\pm}_{m_j}\rangle}$, for $j\geq2$,
is given by Eq.~(\ref{ProbA}).

The overall rate of success is given by 
\begin{equation}
\tilde{P}_{suc}^{tot}(n,q+1) = P_{suc}^{tot}(n,q) + \tilde{P}^{(q+1)}_{suc}.
\label{Ptilde}
\end{equation}
But it is not difficult to show that
\begin{eqnarray}
\tilde{P}_{suc}^{tot}(n,q+1) &=& 
\frac{2(1-n^2)}{(1+n^2)}\left(\sum_{j=2}^q\frac{n^{2^{j-1}}}{(1-n^{2^{\,\!^j}})}
+\frac{n^{2^{q}}}{(1-n^{2^{q}})}\right)\nonumber \\
&=& \frac{2(1-n^2)}{(1+n^2)}\left(\!\sum_{j=2}^{q-1}\frac{n^{2^{j-1}}}{(1-n^{2^{\,\!^j}})}
+\frac{n^{2^{q-1}}}{(1-n^{2^{q-1}})}\right)=\tilde{P}_{suc}^{tot}(n,q). 
\label{qq}
\end{eqnarray}

This is one of the main results in this paper. It states that the overall
probability of success is the same whether we implement $q$ or
$q-1$ teleportations and subsequently correct the remaining unsuccessful 
cases using the strategies of group $1$. 
It does not matter how many teleportations we do, the final probability of success
is the same. 

Equation (\ref{qq}) also allows us to get a simple expression for 
$\tilde{P}_{suc}^{tot}(n,q)$. Indeed, since 
$\tilde{P}_{suc}^{tot}(n,q) =\tilde{P}_{suc}^{tot}(n,q-1)$ we obviously
have $\tilde{P}_{suc}^{tot}(n,q) =\tilde{P}_{suc}^{tot}(n,2)$, which leads
to 
\begin{equation}
\tilde{P}_{suc}^{tot}(n,q) = \frac{2n^2}{1+n^2}.
\label{prob3ch}
\end{equation}

To complete the analysis we now prove that 
this new protocol outperforms the best/third protocol
of \cite{Rig09}.  In other words, we want to show that
$q-1$ teleportations followed by corrections via strategies
of group $1$ is more efficient than $q$ teleportations.
Mathematically, we want to prove that
$P_{suc}^{tot}(n,q)<\tilde{P}_{suc}^{tot}(n,q)$.

First we note that using Eq.~(\ref{Ptilde}) we have
\begin{eqnarray}
P_{suc}^{tot}(n,q)&=& \tilde{P}_{suc}^{tot}(n,q+1) - \tilde{P}^{(q+1)}_{suc}
\nonumber \\
P_{suc}^{tot}(n,q)&=& \tilde{P}_{suc}^{tot}(n,q) - \tilde{P}^{(q+1)}_{suc},
\label{limit1}
\end{eqnarray}
where the last line comes from Eq.~(\ref{qq}). Therefore, since
$\tilde{P}^{(q+1)}_{suc}>0$ we immediately have 
\begin{equation}
P_{suc}^{tot}(n,q) < \tilde{P}_{suc}^{tot}(n,q).
\end{equation}

Also, using Eq.~(\ref{prob3ch}) we can write Eq.~(\ref{limit1}) as follows
\begin{equation}
P_{suc}^{tot}(n,q) =  \frac{2n^2}{1+n^2} - \tilde{P}^{(q+1)}_{suc}.
\end{equation}
Moreover, using Eq.~(\ref{total2}) it is not difficult to see that 
$\lim_{q\to \infty}\tilde{P}^{(q+1)}_{suc}=0$ leading to
\begin{equation}
\lim_{q\to \infty} P_{suc}^{tot}(n,q) =  \frac{2n^2}{1+n^2}.
\end{equation}
In other words, without applying the strategies of group $1$,
only with an infinity number of teleportations we can equal the
efficiency of the new protocol here presented. 

Finally, it is worth mentioning that Eq.~(\ref{prob3ch}) is exactly equal to $P_{suc_1}$, Eq.~(\ref{group1}).
Hence, in a realistic setting one should implement just one teleportation followed by the unitary
correction as described in group $1$. If we keep teleporting the state we will introduce at each teleportation
more and more errors due to imperfect Bell measurements and quantum channels. 
Putting it simply, less is more when it comes to the number of 
teleportations in a real world implementation of the previous protocols.

\section{Direct multiple teleportation in one-dimensional networks}

\begin{figure}[!ht]
\begin{center}
\includegraphics[width=7cm]{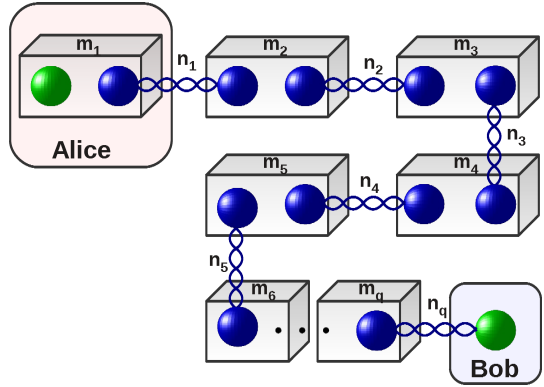}
\end{center}
\caption{\label{fig3}  Now Alice and Bob do not share any entangled state. 
They only share entangled states (one for each) with intermediate parties/stations, where the
other Bell measurements (boxes) are implemented. Bob receives the Bell measurement results from
every station which allows him to properly implement the usual Pauli unitary correction
on the output qubit and other correction strategies as described in the text. 
A protocol is deemed successful only if Bob's output state can be brought to
$\alpha|0\rangle+\beta|1\rangle$ at his place. }
\end{figure}

We now increase the constraint on the resources available to Bob. 
Contrary to the protocols just presented, this time Bob has access 
to only one partially entangled state. This quantum channel is 
not shared with Alice and the goal of the protocol is to deliver 
Alice's input with unity fidelity to Bob after $q$ teleportations. 
The intermediate teleportations are implemented by additional reliable parties 
that inform the result of each Bell measurement to Bob (See Fig. \ref{fig3}). 
With this information Bob knows what Pauli unitary operation to apply
on his qubit as well as whether or not the state at his possession is
exactly $\alpha|0\rangle+\beta|1\rangle$. He also knows when the protocol
fails and how to implement the appropriate corrective strategy as we 
describe in what follows.

The first protocol we present is the one in which $m_j=n_j=n$, i.e.,
we assume generalized Bell measurements at each teleportation with
the matching condition. As already explained in the previous
section, in this scenario the output qubit 
after the $j$-th teleportation changes as  
$(\alpha_{j},\beta_j) \rightarrow (\alpha_{j}, n^2\beta_{j})$
if $\ket{\Phi^+_{m_j}}$ is the result of the generalized Bell
measurement, as $(\alpha_{j},\beta_j) \rightarrow n(\alpha_{j},\beta_{j})$
if it is either $\ket{\Phi^-_{m_j}}$ or $\ket{\Psi^+_{m_j}}$, and
as $(\alpha_{j},\beta_j) \rightarrow (n^2\alpha_{j},\beta_{j})$
if $\ket{\Psi^-_m}$. With this rule in mind a direct count of 
the cases in which we have a complete success after the $q$-th
teleportation leads to the following success rate,
\begin{equation}
P_{suc}^{tot}(n,q) = \binom{2q}{q}\frac{n^{2q}}{(1+n^2)^{2q}},
\end{equation}
with $\binom{a}{b}=a!/((a-b)!b!)$ being the binomial coefficient.

The unsuccessful cases will be given by either $\binom{2q}{q-j}$ unnormalized states
$\alpha|0\rangle+n^{2j}\beta|1\rangle$ or by $\binom{2q}{q-j}$ 
states $n^{2j}\alpha|0\rangle+\beta|1\rangle$, 
$j=1,\ldots, q$. To each one of these cases we add an ancilla and apply
the strategies of group $1$ to correct them. Adding all cases leads to a probability of 
success given by
\begin{equation}
\tilde{P}^{(q+1)}_{suc} = \frac{2}{(1+n^2)^{2q}}
\sum_{j=1}^{q}\binom{2q}{q-j}n^{2(q+j)}.
\end{equation}
Hence, the overall success rate after $q$ teleportations and corrections
via strategies of group $1$ is   
%
\begin{eqnarray}
\tilde{P}_{suc}^{tot}(n,q+1) = P_{suc}^{tot}(n,q) + \tilde{P}^{(q+1)}_{suc}
= \frac{1}{(1+n^{2})^{2q}}\left[\sum_{j=1}^{q}\binom{2q}{q-j}n^{2(q+j)}
+\sum_{j=0}^{q}\binom{2q}{q+j}n^{2(q+j)}\right],
\label{Ptilde2}
\end{eqnarray}
%
where we have used that $\binom{a}{b}=\binom{a}{a-b}$ to arrive at the last 
expression. Note that the last sum starts at zero.

The second protocol we study is simpler and gives a better
result than the previous one.  Now we have standard Bell measurements
$m_j=1$ and quantum channels given by $n_j=n$. 
In this case the output qubit 
after the $j$-th teleportation changes as  
$(\alpha_{j},\beta_j) \rightarrow (\alpha_{j}, n\beta_{j})$
if $\ket{\Phi^{\pm}_{m_j}}$ is the result of a Bell
measurement and  $(\alpha_{j},\beta_j) \rightarrow (n\alpha_{j},\beta_{j})$
if we get $\ket{\Psi^{\pm}_m}$. Since only a balanced sequence with
equal numbers of $\ket{\Phi^{\pm}_m}$ and $\ket{\Psi^{\pm}_m}$ leads
to a direct success, we have $P_{suc}^{tot}(n,q)= 0$ for odd
$q$. For all instances of odd $q$ or for the remaining unsuccessful cases of
even $q$ we apply the appropriate strategies of group $1$. 

Whenever we have an even $q$ we will be dealing with 
$2^q\times \binom{q}{q/2}$ direct successful cases with
the following total probability of direct success,
\begin{equation}
P_{suc}^{tot}(n,q) =
\left\{
\begin{array}{cc}
\binom{q}{q/2}\frac{n^{q}}{(1+n^2)^{q}}, & \mbox{even} \hspace{.1cm} q,\\
0, & \mbox{odd} \hspace{.1cm} q. 
\end{array}
\right.
\label{inter}
\end{equation}

The unsuccessful cases, however, are $2^q\times\binom{q}{j}$ unnormalized states
$\alpha|0\rangle+n^{q-2j}\beta|1\rangle$ or $2^q\times\binom{q}{j}$ 
states $n^{q-2j}\alpha|0\rangle+\beta|1\rangle$, 
$j=0,\ldots, [(q-1)/2]$. Here $[a]$ represents the integer part of the number $a$.
We correct every one of these cases with the strategies of group $1$. Adding all cases 
corrected via strategies of group $1$ leads to a probability of 
success given by
\begin{equation}
\tilde{P}^{(q+1)}_{suc} =
\sum_{j=0}^{[(q-1)/2]}\binom{q}{j}\frac{2n^{2(q-j)}}{(1+n^2)^q}.
\end{equation}
Thus, the overall success rate after $q$ teleportations and corrections
via strategies of group $1$ is   
%
\begin{eqnarray}
\tilde{P}_{suc}^{tot}(n,q+1) 
= P_{suc}^{tot}(n,q) + \tilde{P}^{(q+1)}_{suc}
=P_{suc}^{tot}(n,q) +\hspace{-.3cm} 
\sum_{j=0}^{[(q-1)/2]}\binom{q}{j}\frac{2n^{2(q-j)}}{(1+n^2)^q},
\label{Ptilde3}
\end{eqnarray}
%
with $P_{suc}^{tot}(n,q)$ given by Eq.~(\ref{inter}).

\begin{figure}[!ht]
\begin{center}
\includegraphics[width=7cm]{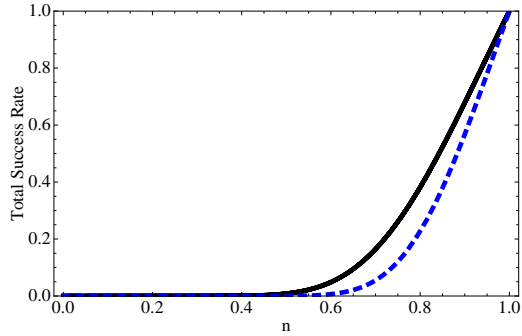}
\caption{\label{fig3a}  Overall probability of success for the
second protocol (black/solid curve), Eq.~(\ref{Ptilde3}), 
and the first protocol (blue/dashed curve), Eq.~(\ref{Ptilde2}), for a network connected
by $q=15$ partially entangled states with $n_j=n$.}
\end{center}
\end{figure}

This second protocol is more efficient than the previous one and in Fig.~\ref{fig3a}
we illustrate this point for the case of $q=15$ successive teleportations.

Moreover, we have checked algebraically up to $q=100$ and numerically for up to $q=10000$ that
Eq.~(\ref{Ptilde3}) is exactly the same as
\begin{equation}
\tilde{P}_{suc}^{tot}(n,q+1) = 1-(f_n^2-g_n^2)\sum_{j=0}^{[(q-1)/2]}f_n^{2j}g_n^{2j}\binom{2j}{j},
\end{equation}
where $f_n=1/\sqrt{1+n^2}$ and $g_n=n/\sqrt{1+n^2}$. This last expression \cite{Per08,Rig09}
is exactly equal to the probability to distill a maximally entangled state between Alice and
Bob out of a network of $q$ identical pure partially entangled states \cite{Per08} connecting them
(See Fig. \ref{fig3}). This is another main result of this paper. It says that we obtain the same
overall probability of success if we work with the direct approach here presented or with 
the indirect (distillation) approach of \cite{Per08,note1}. In the former approach 
the teleported qubit travel through every node/station of the network while in 
the latter if flies directly from Alice to Bob through the distilled maximally 
entangle state.

\section{Going beyond serial Bell states}

We want in this section to present several new partially entangled states
protocol that goes beyond the ``serial circuit'' paradigm of quantum channels
composed of Bell states and that combine, at the same time, the corrective 
strategies described in Sec. \ref{sec2}. 
We modify the previous protocols introducing either
an additional Bell state ``parallel'' to the one already shared between Alice
and Bob or by using other types of quantum channels and measurement basis, 
specifically partially entangled Greenberger-Horne-Zeilinger (GHZ) states
\cite{Gre89,Gor06B}.

\subsection{Two parallel Bell states}

There exist at least three ways in which these sorts of protocols can be built.
The first one consists in measuring the three qubits, the one
to be teleported and the other two coming from half of each channel, 
using a three qubit generalized GHZ basis \cite{Gor06B,Li04} 
(see Fig. \ref{fig4}). The second approach introduces an additional
qubit that interacts with the one to be teleported. We then implement
the traditional two qubit teleportation protocol \cite{Gor06A,Rig05} followed
by corrections to the unsuccessful cases using the strategies of group $1$
(see Fig. \ref{fig5}). In a certain sense, we are masking the original qubit 
to be teleported into a two qubit input state. 
Finally, the third approach also uses this masking strategy
but, instead of performing a two distinct 
generalized Bell measurements as in the second approach, it realizes
a generalized four qubit GHZ measurement involving the two qubit
state to be teleported and the two qubits coming from half of each
quantum channel (see Fig. \ref{fig8}). We now pass to detailing each
one of those three approaches.

\subsubsection{Generalized three qubit GHZ measurement}

In Ref. \cite{Gor06B} many of the steps of the following protocol
were studied in detail and we just briefly review them. The
main changes introduced here, and which we discuss more thoroughly,
are at the final steps of the protocol,
after Alice informs Bob of her measurement outcome. 

\begin{figure}[!ht]
\begin{center}
\includegraphics[width=5cm]{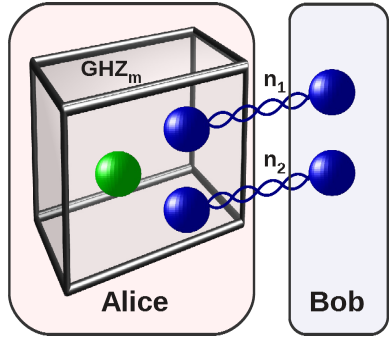}
\end{center}
\caption{\label{fig4}  In this protocol we use two parallel Bell states
$\ket{\Phi^+_{n_1}}$ and $\ket{\Phi^+_{n_2}}$ and measure the three qubits at the left
in the $GHZ_m$ basis. Depending on the measurement outcome at Alice's the state with Bob will be
described by the clean state $\alpha\ket{0}+\beta\ket{1}$ or by a noisy one.
The last step of the protocol consists in applying the strategies of group $1$ of
Sec. \ref{sec2} to these noisy outputs.}
\end{figure}

Looking at Fig. \ref{fig4} the initial state of the system is
\begin{equation}
\ket{\Psi}_{13546} = \ket{\phi}_1\ket{\Phi_{n_1}^+}_{35}\ket{\Phi_{n_2}^+}_{46}, 
\label{ab1}
\end{equation}
with qubit $1$ given by Eq.~(\ref{eq:qubitoriginaldenovo}) and 
the generalized Bell states given by Eq.~(\ref{canaln}) with
$n_1=n_2=n$. Note that we assume both channels have the same entanglement
(matching condition).
We tested some cases with different entanglement and obtained no better performance,
although no general proof was found that the matching condition is 
the optimal choice in this scenario (See Sec. \ref{last}).

Defining the generalized $GHZ_m$ basis as
\begin{eqnarray}
|GHZ^{+}_{m}\rangle = M (|000\rangle + m |111\rangle),&\label{GHZ1}
|GHZ^{-}_{m}\rangle = M (m|000\rangle - |111\rangle),\\
|G^{+}_{m}\rangle = M (|010\rangle + m |101\rangle),&
|G^{-}_{m}\rangle = M (m|010\rangle - |101\rangle),\\
|H^{+}_{m}\rangle = M (|100\rangle + m |011\rangle),&
|H^{-}_{m}\rangle = M (m|100\rangle - |011\rangle),\\
|Z^{+}_{m}\rangle = M (|110\rangle + m |001\rangle),&
|Z^{-}_{m}\rangle = M (m|110\rangle - |001\rangle),
\label{GHZ2}
\end{eqnarray} 
with $m$ real, $0<m<1$, and $M=1/\sqrt{1+m^2}$ we can write Eq.~(\ref{ab1})
as
\begin{eqnarray}
|\Phi\rangle_{13456} & = & MN^2 \left[ 
|GHZ^+\rangle \left( \alpha|00\rangle + mn^2 \beta |11\rangle\right) \right.
+ |GHZ^-\rangle \left( m\alpha|00\rangle -n^2 \beta |11\rangle\right)
\nonumber \\
& & + |Z^+\rangle \left( mn\alpha|01\rangle + n\beta |10\rangle\right)
+ |Z^-\rangle \left( -n\alpha|01\rangle + mn\beta |10\rangle\right)
\nonumber \\
& & 
+ |G^+\rangle \left( n\alpha|10\rangle + mn\beta |01\rangle\right)
+ |G^-\rangle \left( mn\alpha|10\rangle - n\beta |01\rangle\right)
\nonumber \\
& & 
+ |H^+\rangle \left( mn^2\alpha|11\rangle + \beta |00\rangle\right)
\left. + |H^-\rangle \left( -n^2\alpha|11\rangle
+ m\beta |00\rangle\right)\right],
\label{ab2}
\end{eqnarray}
where $N=1/\sqrt{1+n^2}$. In the previous equation we have changed the ordering of the qubits in
a way that now the first three qubits (those with Alice) are the ones that will be measured
in the $GHZ_m$ basis.

We tested the cases where $m=n$ and $m=1$. The latter gave the best overall probability
of success and is the one we detail now. Assuming $m=1$ we immediately see that whenever
Alice measures either $|Z^{\pm}\rangle$ or $|G^{\pm}\rangle$, Bob's output state can be 
brought to 
%
$\alpha\ket{01}+\beta\ket{10}$
if he applies, respectively, the following unitary operations on his two qubits,
$\mathbb{1}\otimes \mathbb{1}$, $\mathbb{1}\otimes \sigma_z$, $\sigma_x\otimes \sigma_x$,
and $\sigma_x\otimes \sigma_{x}\sigma_z$. He knows which unitary operation to apply
because Alice informs him of her measurement outcome.

In order to retrieve the single qubit teleported by Alice, Bob implements 
a CNOT gate on his two qubits
with the first one being the control qubit. After that operation his state is
simply $(\alpha\ket{0}+\beta\ket{1})\ket{1}$ and he finally gets the 
correct teleported state. The total probability of success at this stage of the protocol
(the sum of the probability of each one of these four cases) is
\begin{equation}
P_{suc}^{tot}(n) = \frac{2n^2}{(1+n^2)^2}.
\end{equation}

It is important to note that these four instances of success occur
whether or not we know the entanglement of the channel. We just need
to guarantee that we have two channels with the same degree of entanglement $n$.
This is another main result of the paper and a feature that will appear again in the
other two protocols that follows. Whenever we use two identical generalized Bell states
in parallel we will have some cases of success not depending
on our knowledge of the entanglement of the channels.

We increase the rate of success of this protocol by correcting those four 
cases that do not lead to a direct success using the strategies of group $1$
of Sec. \ref{sec2}. Looking at Eq.~(\ref{ab2}) we note that Bob 
can bring his two qubits after an unsuccessful event  
to one of the following two unnormalized forms, 
$\alpha\ket{00}+n^2\beta\ket{11}$ or $n^2\alpha\ket{00}+\beta\ket{11}$,
after applying the appropriate Pauli unitary operation. Subsequently he
applies a CNOT gate with the first one being the control. The control qubit
can thus be written as either $\alpha\ket{0}+n^2\beta\ket{1}$ or 
$n^2\alpha\ket{0}+\beta\ket{1}$. These states can be corrected adding an
ancilla and applying the techniques of Sec. \ref{sec2}. This leads to
the following probability of correcting all these four instances,
\begin{equation}
\tilde{P}_{suc} = \frac{2n^4}{(1+n^2)^2}, 
\end{equation}
and, hence, to the following overall probability of success
\begin{equation}
\tilde{P}_{suc}^{tot}(n) = P_{suc}^{tot}(n) + \tilde{P}_{suc} = \frac{2n^2}{1+n^2}.
\label{probab}
\end{equation}

\subsubsection{Generalized two qubit teleportation protocol}

The following protocol is an adaptation and improvement 
of the one given in Ref. \cite{Gor06A}. There, a complete probabilistic
strategy on how to teleport an arbitrary two qubit state 
$\alpha\ket{00}+\beta\ket{01}+\gamma\ket{10}+\delta\ket{11}$ with
unity fidelity using two partially entangled states was presented. 
Our goal here, however, is to use that protocol to teleport one qubit.
For that purpose, we interact an ancilla ($\ket{0}$) with Alice's qubit
using a CNOT gate. The qubit to be teleported act as the control qubit. 
After the interaction, the two qubit state with Alice is given by
\begin{equation}
\ket{\phi}_{12}=\alpha\ket{00}+\beta\ket{11}. 
\end{equation}
The qubit to be teleported is embedded in a two qubit state or, equivalently,
``masked'' as a two qubit state. 

\begin{figure}[!ht]
\begin{center}
\includegraphics[width=5cm]{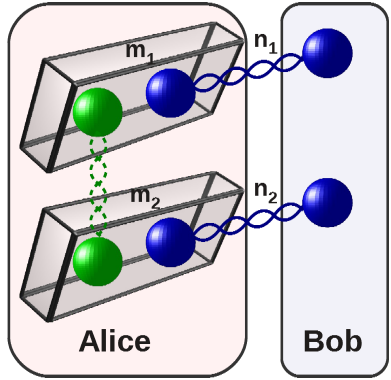}
\end{center}
\caption{\label{fig5}  Here we use two parallel Bell states
$\ket{\Phi^+_{n_1}}$ and $\ket{\Phi^+_{n_2}}$ and measure each pair of qubits with Alice using two $B_m$ basis
(boxes $m_1$ and $m_2$). The qubit to be teleported is embedded in a two qubit state
as described in the text. For certain measurement outcomes Bob's state will be
described by the clean state $\alpha\ket{0}+\beta\ket{1}$. For other measurement
results Bob will need to apply the strategies of group $1$ of
Sec. \ref{sec2} to correct the noisy outputs.}
\end{figure}

Looking at Fig. \ref{fig5} the global state can be written as
\begin{equation}
\ket{\Psi}_{123546} = \ket{\phi}_{12}\ket{\Phi_{n_1}^+}_{35}\ket{\Phi_{n_2}^+}_{46}.
\label{aab1}
\end{equation}

Rewriting the qubits in the order $132456$, where qubits $1$ and
$3$ will be measured in the generalized Bell basis $B_{m_1}$ and
qubits $2$ and $4$ in the basis $B_{m_2}$, we have
%
\begin{eqnarray}
|\Phi\rangle_{132456} & = & M^{2}N^{2}\left[
|\Phi_{m_{1}}^{+}\rangle|\Phi_{m_{2}}^{+}\rangle \left(
\alpha |00\rangle +n^2 \beta|11\rangle\right) \right. 
+ |\Phi_{m_{1}}^{+}\rangle|\Phi_{m_{2}}^{-}\rangle \left(
\alpha |00\rangle -n^2 \beta|11\rangle\right) \nonumber \\
& &+ |\Phi_{m_{1}}^{+}\rangle|\Psi_{m_{2}}^{+}\rangle \left(
n \alpha|01\rangle  +n \beta|10\rangle\right)
 + |\Phi_{m_{1}}^{+}\rangle|\Psi_{m_{2}}^{-}\rangle \left(
n \alpha|01\rangle -n \beta|10\rangle\right) \nonumber \\
& &+ |\Phi_{m_{1}}^{-}\rangle|\Phi_{m_{2}}^{+}\rangle \left(
\alpha |00\rangle -n^2 \beta|11\rangle\right)
+ |\Phi_{m_{1}}^{-}\rangle|\Phi_{m_{2}}^{-}\rangle \left(
\alpha |00\rangle +n^2 \beta|11\rangle\right) \nonumber \\
& & + |\Phi_{m_{1}}^{-}\rangle|\Psi_{m_{2}}^{+}\rangle \left(
n \alpha|01\rangle -n \beta|10\rangle\right) 
+ |\Phi_{m_{1}}^{-}\rangle|\Psi_{m_{2}}^{-}\rangle \left(
n \alpha|01\rangle +n \beta|10\rangle\right) \nonumber \\
& & + |\Psi_{m_{1}}^{+}\rangle|\Phi_{m_{2}}^{+}\rangle \left(
n \alpha|10\rangle  +n\beta|01\rangle\right)
+ |\Psi_{m_{1}}^{+}\rangle|\Phi_{m_{2}}^{-}\rangle \left(
n \alpha|10\rangle  -n \beta|01\rangle\right) \nonumber \\
& & + |\Psi_{m_{1}}^{+}\rangle|\Psi_{m_{2}}^{+}\rangle \left(
n^2 \alpha|11\rangle +\beta |00\rangle\right) 
+ |\Psi_{m_{1}}^{+}\rangle|\Psi_{m_{2}}^{-}\rangle \left(
n^2 \alpha|11\rangle -\beta |00\rangle\right) \nonumber \\
& & + |\Psi_{m_{1}}^{-}\rangle|\Phi_{m_{2}}^{+}\rangle \left(
n \alpha|10\rangle -n \beta|01\rangle\right)  
+ |\Psi_{m_{1}}^{-}\rangle|\Phi_{m_{2}}^{-}\rangle \left(
n \alpha|10\rangle +n \beta|01\rangle\right) \nonumber \\ 
& & + |\Psi_{m_{1}}^{-}\rangle|\Psi_{m_{2}}^{+}\rangle \left(
n^2 \alpha|11\rangle -\beta |00\rangle\right)
\left. + |\Psi_{m_{1}}^{-}\rangle|\Psi_{m_{2}}^{-}\rangle
\left( n^2 \alpha|11\rangle +\beta|00\rangle\right) 
\right], \label{aab2}
\end{eqnarray}
%
where $M = 1/\sqrt{1+m^{2}}$ and $N = 1/\sqrt{1+n^{2}}$. Here we have
assumed $n_1=n_2=n$ and $m_2=m_2=m=1$. We have also tested
other possible strategies, i.e., all possible combinations
of $m_1=1,n,n^2$ and $m_2=1,n,n^2$. Out of these nine possibilities
the best one was $m_1=m_2=1$.

Looking at Eq.~(\ref{aab2}) we see that there are eight possibilities 
that can be brought to $\alpha |00\rangle +\beta|11\rangle$
after appropriate Pauli unitary corrections. For these instances
Bob recovers the original qubit applying a CNOT operation having
the first qubit as the control qubit (unmasking). Note that again these eight
possibilities of direct success happen whether or not we know the 
entanglement content of the channel. All we must have is that both
channels are identical. The eight direct possible outcomes have the same
probability of occurrence given a total of
\begin{equation}
P_{suc}^{tot}(n) = \frac{2n^2}{(1+n^2)^2}.
\end{equation}

The other eight outcomes can be transformed to 
either $\alpha|00\rangle + n^2\beta\ket{11}$ or 
$n^2\alpha|00\rangle + \beta\ket{11}$ by appropriate 
Pauli unitary operations. Bob then applies a 
CNOT gate having the first qubit as the control to obtain
either $\alpha|0\rangle + n^2\beta\ket{1}$ or 
$n^2\alpha|0\rangle + \beta\ket{1}$ as the state describing
the control qubit. After applying the strategies of group $1$
of Sec. \ref{sec2} we have a probability of success equals to
\begin{equation}
\tilde{P}_{suc} = \frac{2n^4}{(1+n^2)^2}. 
\end{equation}
Again, the overall probability of success becomes
\begin{equation}
\tilde{P}_{suc}^{tot}(n) = P_{suc}^{tot}(n) + \tilde{P}_{suc} = \frac{2n^2}{1+n^2}.
\label{probaab}
\end{equation}

\subsubsection{Generalized four qubit GHZ measurement}

\begin{figure}[!ht]
\begin{center}
\includegraphics[width=5cm]{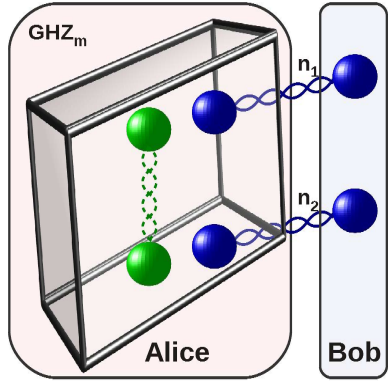}
\end{center}
\caption{\label{fig8}  The quantum channels
are two generalized Bell states ($n_1$ and $n_2$) and the teleported
state is a two qubit system, in which the qubit to be teleported is
embedded (masking technique). Alice realizes a joint four qubit projection
of her qubits onto the generalized four qubit $GHZ_m$ basis. Depending on the
relationship between the values of $n_1, n_2$, and $m$ there exist measurement
results that lead to a perfect teleportation. For the unsuccessful cases
Bob relies on the strategies of group $1$ described in Sec. \ref{sec2} to improve
the success rate.}
\end{figure}

We still keep using a pair of generalized Bell states
and the masking technique of the previous protocol, 
where the qubit to be teleported is embedded in two qubits.
The initial state describing the global state is given by
Eq.~(\ref{aab1}). But now, since Alice will project the four 
qubits at her possession onto the four qubit $GHZ_m$ basis
(see Fig. \ref{fig8}) we rewrite Eq.~(\ref{aab1}) as follows 
\begin{eqnarray}
|\Phi\rangle_{132456} &=& MN^2\left[
|A^+_m\rangle(\alpha|00\rangle + mn^2\beta|11\rangle)\right.
+|A^-_m\rangle (m\alpha|00\rangle -n^2\beta|11\rangle)
\nonumber\\
&&+|B^+_m\rangle(mn\alpha|01\rangle +n\beta|10\rangle)
+|B^-_m\rangle(-n\alpha|01\rangle + mn\beta|10\rangle)
\nonumber\\
&&+|E^+_m\rangle(n\alpha|10\rangle + mn\beta|01\rangle)
+|E^-_m\rangle(mn\alpha|10\rangle -n\beta|01\rangle)
\nonumber\\
&&+|F^+_m\rangle(mn^2\alpha|11\rangle +\beta|00\rangle)
\left.+|F^-_m\rangle(-n^2\alpha|11\rangle +m\beta|00\rangle)
\right], 
\label{aabb2}
\end{eqnarray}
where $\ket{A^{\pm}}$, $\ket{B^{\pm}}$,
$\ket{E^{\pm}}$, and $\ket{F^{\pm}}$ are $8$
of the $16$ states forming the generalized four qubit
$GHZ_m$ basis (see \ref{appendixB}).

From here we can proceed studying three cases, namely,
when we set $m=1, n$, or $n^2$. For $m=1$ we obtained, 
after including all
possible successful cases, i.e., the direct ones and 
those corrected via strategies of group $1$,
the best probability of success. For conciseness, and since
the calculations are very similar to the other cases, 
we only show here in details the best situation ($m=1$).

When $m=1$ Eq.~(\ref{aabb2}) tells us that there are
four possible measurement outcomes that lead directly to 
a perfect teleportation: $|B^{\pm}_m\rangle$ or 
$|E^{\pm}_m\rangle$. After appropriate Pauli unitary operations
followed by a CNOT with the first qubit as the control,
Bob can bring the control qubit to $\alpha\ket{0}+\beta\ket{1}$.
The total probability for a direct success (no need of strategies
of group $1$) is given by
\begin{equation}
P_{suc}^{tot}(n) = \frac{2n^2}{(1+n^2)^2}.
\end{equation}
Note that again these four successful cases can occur without 
our knowing the degree of entanglement of the two channels.
If only suffices to know that they are the same.

The remaining four cases can be transformed
to either $\alpha\ket{00}+n^2\beta\ket{11}$ or 
$n^2\alpha\ket{00}+\beta\ket{11}$ after Bob implements the 
appropriate Pauli rotation. Subsequently, applying the strategies
of group $1$ he gets
\begin{equation}
\tilde{P}_{suc} = \frac{2n^4}{(1+n^2)^2}
\end{equation}
for the probability of correcting the states and
\begin{equation}
\tilde{P}_{suc}^{tot}(n) = P_{suc}^{tot}(n) + \tilde{P}_{suc} = \frac{2n^2}{1+n^2}
\end{equation}
for the overall success rate.

\subsection{Multipartite entangled channel protocols}

The last two protocols we present employ genuine multipartite
partially entangled state as quantum channels. In particular, we
use the four qubit generalized $GHZ_m$ state $\ket{A_m^+}$ as
given in \ref{appendixB}. In the first protocol we
project the qubits with Alice onto the generalized $GHZ_m$ basis
and in the second one onto the generalized Bell basis $B_m$.

\subsubsection{Generalized four qubit GHZ measurement}

It is known that maximally entangled three \cite{Kar98} and four qubit 
\cite{Pat00} GHZ states can be used to teleport one qubit from Alice
to Bob. Here, however, we use a multipartite partially entangled state
and the masking technique together with corrections of unsuccessful cases
using the strategies of group $1$ of Sec. \ref{sec2}.

To start the protocol, we interact the qubit to be teleported with an 
ancilla ($\ket{0}$) via a CNOT gate, where the former acts as the control qubit. 
After the interaction (masking), the two qubit state with Alice is given by
\begin{equation}
\ket{\phi}_{12}=\alpha\ket{00}+\beta\ket{11}. 
\end{equation}

The total state can thus be written as
\begin{equation}
\ket{\Psi}_{123546} = \ket{\phi}_{12}\ket{A^+_n}_{3546}.
\label{aaaa1}
\end{equation}

\begin{figure}[!ht]
\begin{center}
\includegraphics[width=5cm]{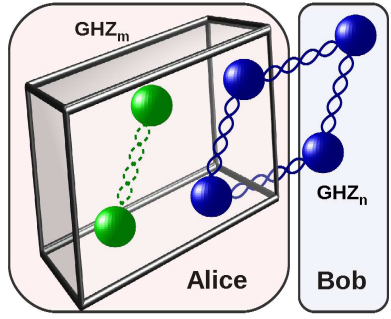}
\end{center}
\caption{\label{fig6}  The quantum channel
is the generalized four qubit $GHZ_n$ state $\ket{A^+_n}$ and the teleported
state is a two qubit system, where the qubit to be teleported is
masked. Alice realizes a joint four qubit projection
of her qubits onto the generalized four qubit $GHZ_m$ basis.}
\end{figure}

Rewriting the state changing the order of the qubits to $132456$ we have
\begin{eqnarray}
|\Phi\rangle_{132456} & = & MN\left[|A^+_m\rangle(\alpha|00\rangle +mn\beta|11\rangle)
\right.
+|A^-_m\rangle(m\alpha|00\rangle-n\beta|11\rangle)
\nonumber\\
&&+|F^+_m\rangle(mn\alpha|11\rangle+\beta|00\rangle)
\left.+|F^-_m\rangle(-n\alpha|11\rangle+m\beta|00\rangle)\right],
\label{aaaa2}
\end{eqnarray}
where $M = 1/\sqrt{1+m^{2}}$ and $N = 1/\sqrt{1+n^{2}}$. Here we have
assumed $n_1=n_2=n$ and $m_1=m_2=m$. 
In this way of writing the order of the qubits, 
qubits $1$ and  $3$ (top left of Fig. \ref{fig6}) and $2$ and $4$  
(bottom left of Fig. \ref{fig6}) will be jointly measured in the generalized 
four qubit $GHZ_{m}$ basis.

When $m=1$ there is no direct success ($P_{suc}^{tot}(n)=0$).
However Bob, after being informed of Alice's measurement result,  
can locally (Pauli rotations) bring his state to either 
$\alpha|00\rangle +n\beta|11\rangle$ or $n\alpha|00\rangle +\beta|11\rangle$.
After a CNOT gate with the first qubit as the control qubit Bob gets either
$\alpha|0\rangle +n\beta|1\rangle$ or $n\alpha|0\rangle +\beta|1\rangle$
for the former, which can be corrected using the strategies of group $1$
given in Sec. \ref{sec2}. The probability of a successful correction
for each one of the four possible measurement results of Alice are the same and equal to 
$n^2/(2(1+n^2))$. Hence, the overall probability of success in this case
becomes
\begin{equation}
\tilde{P}_{suc}^{tot}(n) = \frac{2n^2}{1+n^2}.
\label{equal}
\end{equation}

When $m=n$ two possible Alice's measurement outcomes, $|A^-_m\rangle$ and
$|F^-_m\rangle$, give a direct success 
(no need of strategies of group $1$, just a CNOT), 
leading to $P_{suc}^{tot}(n)=2n^2/(1+n^2)^2$.
The unsuccessful cases can be transformed to
either $\alpha|00\rangle +n^2\beta|11\rangle$ or $n^2\alpha|00\rangle +\beta|11\rangle$
by Pauli rotations and the control qubit to 
$\alpha|0\rangle +n^2\beta|1\rangle$ or $n^2\alpha|0\rangle +\beta|1\rangle$
after a CNOT gate. The last step consists in applying the strategies
of group $1$ giving a correction rate of $\tilde{P}_{suc}=2n^4/(1+n^2)^2$ and
consequently an overall probability of success equals to Eq.~(\ref{equal}).

\subsubsection{Double generalized Bell measurements}

\begin{figure}[!ht]
\begin{center}
\includegraphics[width=5cm]{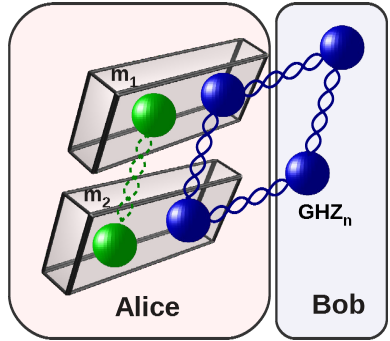}
\end{center}
\caption{\label{fig7}  Again
the quantum channel is given by the $GHZ_n$ state $\ket{A^+_n}$ and the teleported
system is a two qubit system carrying the information of the single
qubit that Alice wants to deliver to Bob (masking technique).  Now, however,
Alice realizes two generalized Bell measurements (boxes $m_1$ and $m_2$).}
\end{figure}

The initial steps of this protocol is identical to the previous one. 
The difference comes when Alice measures her four qubits. Now
qubits $1$ and $3$ are projected onto the basis $B_{m_1}$ and
qubits $2$ and $4$ onto $B_{m_2}$ (see Fig. \ref{fig7}). 

With this information we 
can rewrite the initial state as  
\begin{eqnarray}
|\Phi\rangle_{132456} & = & M^2N
\left[
|\Phi^+_m\rangle|\Phi^+_m\rangle(\alpha|00\rangle+m^2n\beta|11\rangle)
\right.
+|\Phi^+_m\rangle|\Phi^-_m\rangle(m\alpha|00\rangle-mn\beta|11\rangle)
\nonumber\\
&&+|\Phi^-_m\rangle|\Phi^+_m\rangle(m\alpha|00\rangle-mn\beta|11\rangle)
+|\Phi^-_m\rangle|\Phi^-_m\rangle(m^2\alpha|00\rangle+n\beta|11\rangle)
\nonumber\\
&&+|\Psi^+_m\rangle|\Psi^+_m\rangle(n\alpha|11\rangle+m^2\beta|00\rangle)
+|\Psi^+_m\rangle|\Psi^-_m\rangle(mn\alpha|11\rangle-m\beta|00\rangle)
\nonumber\\
&&+|\Psi^-_m\rangle|\Psi^+_m\rangle(mn\alpha|11\rangle-m\beta|00\rangle)
\left.+|\Psi^-_m\rangle|\Psi^-_m\rangle(m^2n\alpha|11\rangle+\beta|00\rangle)
\right],
\label{bbbb2}
\end{eqnarray}
where we have assumed that $m_1=m_2=m$.

We have analyzed three situations, namely,
$m=1$, $m=n$, and $m=n^2$. The cases $m=n$ and $m=n^2$ are 
less efficient than the first one 
and will not be detailed below.

For $m=1$ there is no direct success ($P_{suc}^{tot}(n)=0$) and
Bob must implement the strategies of group $1$ to all $8$ possibilities.
With the knowledge of Alice's measurement result he can Pauli rotate his state to either 
$\alpha|00\rangle +n\beta|11\rangle$ or $n\alpha|00\rangle +\beta|11\rangle$.
After a CNOT gate with the first qubit as the control he obtains either
$\alpha|0\rangle +n\beta|1\rangle$ or $n\alpha|0\rangle +\beta|1\rangle$
for the former. These states can be corrected using the strategies of group $1$
leading to a total probability of success 
\begin{equation}
\tilde{P}_{suc}^{tot}(n) = \frac{2n^2}{1+n^2}.
\end{equation}

\section{Discussion of the results}
\label{last}

For all protocols here presented where Alice and Bob share 
one partially entangled Bell state $\ket{\Phi^+_n}=(\ket{00}+n\ket{11})/\sqrt{1+n^2}$, 
there always exists a configuration where we can achieve a probability
of success given by the `magic' number $P_{suc}=2n^2/(1+n^2)$ after combining
the three major techniques described in Sec. \ref{sec2}. 
This magic number can be understood using the  
important result given in \cite{Vid99}. 

Indeed, in \cite{Vid99}
it was shown that if Alice and Bob share a partially entangled
state as listed above and are restricted to local operations and 
classical communication (LOCC),
the optimal probability to convert one copy of $\ket{\Phi^+_n}$ 
to a maximally entangled state is exactly $P_{suc}$. 
This result implies that there is no other protocol or strategy to increase 
the chances of Alice and Bob's sharing a maximally entangled state if they are restricted
to LOCC and only one state $\ket{\Phi^+_n}$. 

Now, if we could come up with
a direct teleportation protocol using  $\ket{\Phi^+_n}$ as the quantum channel and with
a probability of success greater than $P_{suc}$, we could use it to teleport from
Alice to Bob a qubit maximally entangled with another one at Alice's (entanglement
swapping). But this would be exactly a LOCC protocol that 
makes Alice and Bob share a maximally entangled Bell state
with a higher success rate than the optimal one, contradicting thus the theorem of \cite{Vid99}.
Hence, there is no direct teleportation protocol with only one copy of $\ket{\Phi_n^+}$
shared between Alice and Bob that has a probability of success greater than 
$P_{suc}$.  

However, if Alice and Bob share two partially entangled Bell states 
$\ket{\Phi_n^+}$, the results of \cite{Vid99} tell us that Alice and
Bob can distill one maximally entangled Bell state out of the two $\ket{\Phi_n^+}$
with at least the following
probability of success, $Q_{suc}=P_{suc}+(1-P_{suc})P_{suc}=4n^2/(1+n^2)^2$,
with $Q_{suc}\geq P_{suc}$ and the equality occurring only at the trivial
cases $n=0$ and $n=1$. Therefore, with two partially entangled Bell states the argument of
the previous paragraph cannot be applied and there exists, apparently, no impediment
to the construction of a direct teleportation protocol that has an
efficiency rate greater than $P_{suc}$ and equals to $Q_{suc}$. In spite of this,
and after many attempts described in the previous sections, 
we were not able to build a two channel direct teleportation protocol 
with a probability of success given by $Q_{suc}$. We believe, though,
that there might be a way to build one or, less likely though much more 
interesting, there may be an additional restriction dictated by the laws of 
quantum mechanics that forbids such protocols.

\section{Conclusions}

We have studied how partially entangled pure states can
be used to \textit{directly} teleport a qubit from Alice to Bob
in a variety of configurations. (By direct teleportation we mean 
that no distillation techniques are allowed.) We first
reviewed the three standard main techniques
where one can achieve such a feat and then we compared the 
performances of these three techniques under realistic settings, 
where detection and measurement inefficiencies are taken into account.

Subsequently, we have investigated how the combination of these
techniques could lead to the improvement of multiple teleportation protocols,
where the qubit with Alice are repeatedly teleported along a linear chain of 
many partially entangled states. We analyzed two different scenarios here,
one in which Bob has access to all partially entangled states (Fig. \ref{fig2}) and another 
where Alice and Bob share no entangled state (Fig. \ref{fig3}). 
For both situations we presented protocols that achieve the best efficiency rates to date.
Also, in the second situation our protocol equals the best distillation strategy known
in the literature.

We then moved to the study of protocols that either employ two 
partially entangled Bell states in parallel (Figs. \ref{fig4}, \ref{fig5}, and 
\ref{fig8}) or multipartite partially entangled states (Figs. \ref{fig6} and
\ref{fig7}) to teleport one qubit. In order to build these protocols,
we introduced the `masking' technique, where one qubit is embedded in a two
qubit state. For some of these protocols, we showed that Alice can teleport her qubit to
Bob without even knowing the degree of the entanglement of the two
generalized Bell states. All she needs to know is that the two channels are
the same. 
Finally, we proved that for some of these protocols their rate of success are optimal
and calculated an upper bound for those we could not explicitly build 
the optimal strategies.

\section*{Acknowledgments}
RF thanks CAPES (Brazilian Agency for the Improvement of Personnel of Higher Education) for
partially funding this research. GR thanks CNPq (Brazilian National Science Foundation), 
FAPESP (State of S\~ao Paulo Science Foundation), 
and CNPq/FAPESP for financial support through the 
National Institute of Science and Technology for Quantum Information (INCT-IQ).

\appendix

\section{General analysis for two successive teleportations}
\label{appendixA}

Here we assume different generalized Bell measurements and different entangled
states at each teleportation. For the first round we have $B_m$ labeled by $m_a$ and
the entangled state by $n_a$ while for the second teleportation $m_b$ and $n_b$, respectively.
Also, qubit $1$ and $2$ are with Alice. Qubit $1$ is the input state that will be teleported.
Qubit $2$ is entangled with qubit $3$ at Bob's. The second teleportation is implemented by Bob,
who also have the entangled qubits $4$ and $5$. Qubit $5$ is the output, i.e., the final 
recipient of the teleported state. In what follows we drop the subscripts $1$ to $5$ at each
corresponding ket in order to simplify the notation. 

Before the first teleportation the global state (Alice's input state to be teleported plus the 
entangled channel) is given by Eq.~(\ref{eq:estgeralindiano1}) with $n\rightarrow n_a$ and
$m\rightarrow m_a$. Alice's probability to measure a particular generalized Bell state is given by
Eqs.~(\ref{p2})-(\ref{p4}) and, depending on the measurement result, Bob's qubit at the end
of the first teleportation is given by one of the four states in Eqs.~(\ref{phi2})-(\ref{phi4}). 

In order to analyze the second teleportation, we have to repeat the previous
calculations for each one of the four possible outcomes of the first teleportation.
This will lead to a total of $4\times 4$ possibilities. We can proceed with this calculation
by noting that for a given basis $B_{m_j}$ and channel $|\Phi_{n_j}^+\rangle$ the output 
state with Bob at the end of the teleportation protocol 
(see Eqs.~(\ref{phi2})-(\ref{phi4})), conditioned on Alice's measurement outcome, 
can be summarized as listed in Tab. \ref{apptable1}.
\begin{table}[!htb]
\caption{\label{apptable1}
Unnormalized state with Bob after the $j$-th teleportation. $\alpha_{j-1}$ and $\beta_{j-1}$
are the coefficients of the input state.}
\begin{center}
\begin{tabular}{cc}
Alice's measurement result  & State after $j$-th teleportation \\ \hline 
$|\Phi^+_{m_j}\rangle$ & $
\alpha_{j-1}|0\rangle + m_jn_j\beta_{j-1}|1\rangle
$ \\
$|\Phi^-_{m_j}\rangle$ & $
m_j \alpha_{j-1}|0\rangle + n_j\beta_{j-1}|1\rangle
$ \\
$|\Psi^+_{m_j}\rangle$ & $
n_j\alpha_{j-1}|0\rangle + m_j \beta_{j-1}|1\rangle
$ \\
$|\Psi^-_{m_j}\rangle$ & $
m_j n_j \alpha_{j-1}|0\rangle + \beta_{j-1}|1\rangle
$ \\ \hline
\end{tabular}
\end{center}
\end{table}   

Hence, using Tab. \ref{apptable1} it is not difficult to see that after two 
teleportations we have the $16$ possibilities listed in Tab. \ref{apptable2}
with the respective probabilities. 

\begin{table}[!htb]
\caption{\label{apptable2}
Unnormalized state with Bob (second column) after two teleportations. 
Column one shows the possible measurement outcomes for Alice (first teleportation)
and for Bob (second teleportation). The third column gives the respective 
probability for the outcome listed in the first column.}
\begin{center}
\begin{tabular}{ccc}
Measurements  & Final state & Probability \\ \hline 
$|\Phi^+_{m_a}\rangle|\Phi^+_{m_b}\rangle$ & $\alpha|0\rangle + m_am_bn_an_b\beta|1\rangle$ &$\xi_1(\alpha,\beta)^2$\\
$|\Phi^+_{m_a}\rangle|\Phi^-_{m_b}\rangle$ & $m_b\alpha|0\rangle + m_an_an_b\beta|1\rangle$ &$\xi_2(\alpha,\beta)^2$\\
$|\Phi^+_{m_a}\rangle|\Psi^+_{m_b}\rangle$ & $n_b\alpha|0\rangle + m_am_bn_a\beta|1\rangle$ &$\xi_3(\alpha,\beta)^2$\\
$|\Phi^+_{m_a}\rangle|\Psi^-_{m_b}\rangle$ & $m_bn_b\alpha|0\rangle + m_an_a\beta|1\rangle$ &$\xi_4(\alpha,\beta)^2$\\ \hline
$|\Phi^-_{m_a}\rangle|\Phi^+_{m_b}\rangle$ & $m_a\alpha|0\rangle + m_bn_an_b\beta|1\rangle$ &$\xi_5(\alpha,\beta)^2$\\
$|\Phi^-_{m_a}\rangle|\Phi^-_{m_b}\rangle$ & $m_am_b\alpha|0\rangle + n_an_b\beta|1\rangle$ &$\xi_6(\alpha,\beta)^2$\\
$|\Phi^-_{m_a}\rangle|\Psi^+_{m_b}\rangle$ & $m_an_b\alpha|0\rangle + m_bn_a\beta|1\rangle$ &$\xi_7(\alpha,\beta)^2$\\
$|\Phi^-_{m_a}\rangle|\Psi^-_{m_b}\rangle$ & $m_am_bn_b\alpha|0\rangle + n_a\beta|1\rangle$ &$\xi_8(\alpha,\beta)^2$\\ \hline
$|\Psi^+_{m_a}\rangle|\Phi^+_{m_b}\rangle$ & $n_a\alpha|0\rangle + m_am_bn_b\beta|1\rangle$ &$\xi_8(\beta,\alpha)^2$\\
$|\Psi^+_{m_a}\rangle|\Phi^-_{m_b}\rangle$ & $m_bn_a\alpha|0\rangle + m_an_b\beta|1\rangle$ &$\xi_7(\beta,\alpha)^2$\\
$|\Psi^+_{m_a}\rangle|\Psi^+_{m_b}\rangle$ & $n_an_b\alpha|0\rangle + m_am_b\beta|1\rangle$ &$\xi_6(\beta,\alpha)^2$\\
$|\Psi^+_{m_a}\rangle|\Psi^-_{m_b}\rangle$ & $m_bn_an_b\alpha|0\rangle + m_a\beta|1\rangle$ &$\xi_5(\beta,\alpha)^2$\\ \hline
$|\Psi^-_{m_a}\rangle|\Phi^+_{m_b}\rangle$ & $m_an_a\alpha|0\rangle + m_bn_b\beta|1\rangle$ &$\xi_4(\beta,\alpha)^2$\\
$|\Psi^-_{m_a}\rangle|\Phi^-_{m_b}\rangle$ & $m_am_bn_a\alpha|0\rangle + n_b\beta|1\rangle$ &$\xi_3(\beta,\alpha)^2$\\
$|\Psi^-_{m_a}\rangle|\Psi^+_{m_b}\rangle$ & $m_an_an_b\alpha|0\rangle + m_b\beta|1\rangle$ &$\xi_2(\beta,\alpha)^2$\\
$|\Psi^-_{m_a}\rangle|\Psi^-_{m_b}\rangle$ & $m_am_bn_an_b\alpha|0\rangle + \beta|1\rangle$ &$\xi_1(\beta,\alpha)^2$
\\ \hline
\end{tabular}
\end{center}
\end{table}   

In Tab. \ref{apptable2} we have 
\begin{eqnarray}
\xi_1(\alpha,\beta)=P_{ab} \sqrt{|\alpha|^2 + m_a^2 m_b^2n_a^2n_b^2|\beta|^2 },\,\,\,
\xi_2(\alpha,\beta)=P_{ab} \sqrt{m_b^2|\alpha|^2 + m_a^2n_a^2n_b^2|\beta|^2 },\\
\xi_3(\alpha,\beta)=P_{ab} \sqrt{n_b^2|\alpha|^2 + m_a^2 m_b^2n_a^2|\beta|^2 },\,\,\,
\xi_4(\alpha,\beta)=P_{ab} \sqrt{m_b^2n_b^2|\alpha|^2 + m_a^2n_a^2|\beta|^2 },\\
\xi_5(\alpha,\beta)=P_{ab} \sqrt{m_a^2|\alpha|^2 + m_b^2n_a^2n_b^2|\beta|^2 },\,\,\,
\xi_6(\alpha,\beta)=P_{ab} \sqrt{m_a^2m_b^2|\alpha|^2 + n_a^2n_b^2|\beta|^2 },\\
\xi_7(\alpha,\beta)=P_{ab} \sqrt{m_a^2n_b^2|\alpha|^2 + m_b^2n_a^2|\beta|^2 },\,\,\,
\xi_8(\alpha,\beta)=P_{ab} \sqrt{m_a^2m_b^2n_b^2|\alpha|^2 + n_a^2|\beta|^2 },
\end{eqnarray}
where $M_j=(1+m_j^2)^{-1/2}$, $N_j=(1+n_j^2)^{-1/2}$, $j=a,b$, and
$P_{ab}=M_aM_bN_aN_b$.

Looking at Tab. \ref{apptable2} we see that for 
$m_a=m_b=1$ and $n_a=n_b=n$ the protocol is successful whenever one obtains
one of the following eight possible measurement outcomes: 
$|\Phi^{\pm}_{m_a}\rangle_{12}|\Psi^{\pm}_{m_b}\rangle_{34}$ and 
$|\Psi^{\pm}_{m_a}\rangle_{12}|\Phi^{\pm}_{m_b}\rangle_{34}$. Indeed, in such
a case the final unnormalized state with Bob is $n(\alpha|0\rangle+\beta|1\rangle)$ and
the total probability of success is $P_{suc_3}$, Eq.~(\ref{group3}). 

\section{The generalized four qubit $GHZ_m$ basis}
\label{appendixB}

The $16$ generalized four qubit $GHZ_m$ states can all be generated from the first
one, $\ket{A^+_m}$, by appropriate application of Pauli operations at each one of the
four qubits. This leads to the following basis
\begin{eqnarray*}
|A^+_m\rangle=M(|0000\rangle +m|1111\rangle),&\,\,\,
|A^-_m\rangle=M(m|0000\rangle-|1111\rangle),&\,\,\,
|B^+_m\rangle=M(|1110\rangle +m|0001\rangle),\\ 
|B^-_m\rangle=M(m|1110\rangle -|0001\rangle),&\,\,\,
|C^+_m\rangle=M(|0010\rangle+m|1101\rangle),&\,\,\,
|C^-_m\rangle=M(m|0010\rangle -|1101\rangle),\\
|D^+_m\rangle=M(|1100\rangle +m|0011\rangle),&\,\,\,
|D^-_m\rangle=M(m|1100\rangle -|0011\rangle),&\,\,\,
|E^+_m\rangle=M(|0100\rangle+m|1011\rangle),\\
|E^-_m\rangle=M(m|0100\rangle -|1011\rangle),&\,\,\,
|F^+_m\rangle=M(|1010\rangle +m|0101\rangle),&\,\,\,
|F^-_m\rangle=M(m|1010\rangle -|0101\rangle),\\
|J^+_m\rangle=M(|0110\rangle +m|1001\rangle),&\,\,\,
|J^-_m\rangle=M(m|0110\rangle -|1001\rangle),&\,\,\,
|K^+_m\rangle=M(|1000\rangle +m|0111\rangle),\\
|K^-_m\rangle=M(m|1000\rangle -|0111\rangle),
\end{eqnarray*}
where $0<m<1$ and $M=1/\sqrt{1+m^2}$.






\begin{thebibliography}{99}

\bibitem{sch35} E. Schr\"odinger, Proc. Camb. Phil. Soc. 31 (1935) 555.

\bibitem{epr} A. Einstein, B. Podolsky, and N. Rosen, Phys. Rev. 47 (1935) 777.

\bibitem{bohr} N. Bohr, Phys. Rev. 48 (1935) 696.

\bibitem{bel64} J. S. Bell, Physica 1 (1964) 195.

\bibitem{livrodonielsen} M. A. Nielsen and  I. L. Chuang, Quantum Computation and
Quantum Information, Cambridge University Press, Cambridge, 2000.

\bibitem{livrodozeilinger}  D. Bouwmeester, A. K. Ekert, and A. Zeilinger (Eds.),
The Physics of Quantum Information, Springer-Verlag, Berlin, 2000.

\bibitem{eke91} A. K. Ekert, Phys. Rev. Lett. 67 (1991) 661.

\bibitem{ben92} C. H. Bennett and S. J. Wiesner, Phys. Rev. Lett. 69 (1992) 2881.

\bibitem{ben93} C. H. Bennett, G. Brassard, C. Crepeau, R. Jozsa, A. Peres, and
W.K. Wootters, Phys. Rev. Lett. 70 (1993) 1895.

\bibitem{bra98} S. L. Braunstein and H. J. Kimble, Phys. Rev. Lett. 80 (1998) 869.

\bibitem{ben96} C. H. Bennett, G. Brassard, S. Popescu, B. Schumacher,
J. A. Smolin, and W. K. Wootters, Phys. Rev. Lett. 76 (1996) 722. 

\bibitem{Guo00} W.-Li Li, C.-Feng Li, and G.-C. Guo, Phys. Rev. A
61 (2000) 034301.

\bibitem{Agr02} P. Agrawal and A. K. Pati, Phys. Lett. A 305
(2002) 12.

\bibitem{Gor06A} G. Gordon and G. Rigolin, Phys. Rev. A 73
(2006) 042309.

\bibitem{Gor06B} G. Gordon and G. Rigolin, Phys. Rev. A 73 (2006) 062316.

\bibitem{Gor07} G. Gordon and G. Rigolin, Eur. Phys. J. D 45 (2007) 347.

\bibitem{Gor10} G. Gordon and G. Rigolin, Opt. Commun. 283 (2010) 184. 

\bibitem{Mod08} J. Mod{\l}awska and A. Grudka, Phys. Rev. Lett.
100 (2008) 110503.

\bibitem{Rig09} G. Rigolin, J. Phys. B: At. Mol. Opt. Phys. 
42 (2009) 235504.

\bibitem{Agr06} P. Agrawal and A. Pati, Phys. Rev. A 74 (2006) 062320.

\bibitem{Per08} S. Perseguers, J. I. Cirac, A. Ac\'{\i}n, M. Lewenstein, and J. Wehr,
Phys. Rev. A 77 (2008) 022308.

\bibitem{note1} In \cite{Rig09} $q$ teleportations means $q-1$ repeaters in \cite{Per08}. Hence,
Fig. $7$ of \cite{Rig09} does not accurately compare the same thing. For a fixed number of
teleportations, the approach of \cite{Rig09} is not more efficient than the optimal
one of \cite{Per08}. Here,
however, we improve the protocols of \cite{Rig09} using the strategies of group $1$.
This leads to a new protocol that does equal in efficiency the best  
1D network distillation protocol in \cite{Per08}.

\bibitem{Gre89} D. M. Greenberger, M. A. Horne, and A. Zeilinger,
Bell's theorem, Quantum Theory, and Conceptions of the Universe, 
Kluwer Academics, Dordrecht, 1989.
  
\bibitem{Li04}  Y. Li, K. Zhang, and K. Peng, Phys. Lett. A 324 
(2004) 420.

\bibitem{Rig05} G. Rigolin, Phys. Rev. A 71 (2005) 032303. 

\bibitem{Kar98} A. Karlsson and M. Bourennane, Phys. Rev. A 58 (1998) 4394.

\bibitem{Pat00} A. K. Pati, Phys. Rev. A 61 (2000) 022308.

\bibitem{Vid99} G. Vidal, Phys. Rev. Lett. 83 (1999) 1046.


\end{thebibliography}







\end{document}